\documentclass[10pt,a4paper]{iopart}
\usepackage{iopams}  
\usepackage{times}
\usepackage[varg]{txfonts}
\usepackage{graphicx}	
\usepackage{float}
\usepackage{bm}%
\usepackage{wrapfig}
\usepackage{color}
\usepackage{url}
\usepackage{calc,subfigure}

\usepackage[colorlinks, pdfborder={0 0 0}]{hyperref}
\definecolor{LinkColor}{rgb}{0.75, 0, 0}
\definecolor{CiteColor}{rgb}{0, 0.5, 0.5}
\definecolor{UrlColor}{rgb}{0, 0, 0.75}
\hypersetup{linkcolor=LinkColor}
\hypersetup{citecolor=CiteColor}
\hypersetup{urlcolor=UrlColor}

\newcommand{\Dl}{D_\ell}
\newcommand{\zl}{z_\ell}
\newcommand{\zs}{z_s}
\newcommand{\Ds}{D_s}
\newcommand{\Dls}{D_{\ell s}}
\newcommand{\vOmega}{\vec{\Omega}}
\newcommand{\Tobs}{T_\mathrm{obs}}

\begin{document}

\title[Lensing Cosmography]{Strong-lensing cosmography using third-generation gravitational-wave detectors}

\author{Souvik~Jana$^1$, Shasvath~J~Kapadia$^{2,1}$, Tejaswi~Venumadhav$^{3,1}$, Surhud~More$^{2}$, Parameswaran~Ajith$^{1,4}\footnote{Author to whom any correspondence should be addressed.}$}
\address{$^1$ International Centre for Theoretical Sciences, Tata Institute of Fundamental Research, Bangalore 560089, India}
\address{$^2$ The Inter-University Centre for Astronomy and Astrophysics, Post Bag 4, Ganeshkhind, Pune 411007, India}
\address{$^3$ Department of Physics, University of California at Santa Barbara, Santa Barbara, CA 93106, USA}
\address{$^4$ Canadian Institute for Advanced Research, CIFAR Azrieli Global Scholar, MaRS Centre, West Tower, 661 University Ave., Suite 505, Toronto, ON M5G 1M1, Canada}
\ead{ajith@icts.res.in}

\vspace{10pt}
\begin{indented}
\item[]\today
\end{indented}

\begin{abstract}
We present a detailed exposition of a statistical method for estimating cosmological parameters from the observation of a large number of strongly lensed binary-black-hole (BBH) mergers observable by next (third) generation (XG) gravitational-wave (GW) detectors. This method, first presented in Jana (2023
Phys. Rev. Lett. 130 261401), compares the observed number of strongly lensed GW events and their time delay distribution (between lensed images) with observed events to infer cosmological parameters. We show that the precision of the estimation of the cosmological parameters does not have a strong dependance on the assumed BBH redshift distribution model. Using the large number of unlensed mergers, XG detectors are expected to measure the BBH redshift distribution with sufficient precision for the cosmological inference. However, a biased inference of the BBH redshift distribution will bias the estimation of cosmological parameters. An incorrect model for the distribution of lens properties can also lead to a biased cosmological inference. However, Bayesian model selection can assist in selecting the right model from a set of available parametric models for the lens distribution. We also present a way to incorporate the effect of contamination in the data due to the limited efficiency of lensing identification methods, so that it will not bias the cosmological inference. 
\end{abstract}

\section{Introduction}

The next (third) generation (XG) of gravitational-wave (GW) detectors will observe a large number of compact binary mergers out to large distances. The expected detection rates of binary-black-hole (BBH) and binary-neutron-star (BNS) mergers are $\sim 10^5 - 10^6$ per year \cite{Maggiore_2020,evans2023ce}. These detectors will observe BBH mergers out to redshifts as large as  $z \sim 100$ \cite{Hall_2019}, providing new avenues to probe cosmology that complement other observations. For example, any black holes (BHs) observed at very high redshifts ($z \gtrsim 10$) are unlikely to be of stellar origin as this is before the epoch of star formation. Hence any observation of BBHs at such redshifts will  provide a strong hint on the existence of primordial BHs, which will also have implications on our understanding of dark matter.  

GW observations of compact binary mergers will allow us to measure the luminosity distance $d_L$ accurately without using any distance ladders, as these objects are absolutely calibrated \emph{standard sirens}~\cite{SchutzNat1986, Holz_Hughes_2005}. An electromagnetic (EM) counterpart of the merger will enable us to measure their redshifts, thus allowing us to populate the Hubble diagram. Due to the limited horizon of EM telescopes such measurements are likely to be possible only at lower redshifts ({$z \lesssim 0.5$}). Thus, such measurements will primarily track the Hubble law $d_L(z)\simeq H_0^{-1} z$, thus providing precise measurement of the Hubble constant. The systematic errors of such GW-based measurements are much better understood than, say, that of Type 1a supernovae. Hence, such observations will greatly aid resolving the current Hubble tension~\cite{Riess2020,Hu_2023}.  

At high redshifts ({$z \gtrsim 0.5$}), it is difficult to observe EM counterparts of compact binary mergers. Additionally, BBH mergers, which can be seen out to very high redshifts, are generally not expected to produce EM counterparts. However, even in the absence of EM counterparts, statistical correlation of the mergers with large scale structure will provide some way of inferring cosmological parameters~\cite{Fishbach_2019, Calore_2020, mukherjee2022crosscorrelating,Gair_2023, Borghi_2024}. Here the Hubble diagram will contain imprints of other cosmological parameters such as the dark matter energy density and the equation of state of the dark energy. Thus they will provide strong tests of cosmological models, such as the currently favoured $\Lambda$CDM model~\cite{Planck_2018}. 

We are unable to measure cosmological redshifts from BBH observations because the masses are degenerate with redshift in the GW signal, thanks to the absence of a preferred mass scale in GR. However, non-gravity physics in compact objects and their progenitors could introduce particular mass scales in compact binaries. For example, neutron stars have a maximum mass that depends on the nuclear equation of state~\cite{Lattimer_2021} and the BH mass distribution can have some features due to pair instability supernovae~\cite{Woosley2007,Belczynski_2016a, Woosley_2017, Heger_2002}. If such features are well understood (see e.g.~\cite{Mastrogiovanni:2021wsd,Mukherjee:2021rtw} for some caveats), they will enable redshifts measurements from compact binaries purely using GW observations~\cite{Messenger:2011gi,Li:2013via,Messenger:2013fya, Ezquiaga_2021_PISN, Ezquiaga_2022,chen2024cosmography}. Such \emph{spectral sirens} will also enable GW cosmography. 

GW standard sirens can also probe another signature of the dark energy sector that is not accessible to EM observations. A generic modified gravity theory induces modifications in the evolution of the cosmological background and perturbations, with respect to the standard model of cosmology. In some modified theories,  GWs propagate at the speed of light but their amplitude will decrease differently from general relativity (GR). Consequently, the luminosity distance estimated from the GW standard sirens would differ from that estimated from their EM counterparts. The next generation detectors will be able search for this deviation, probing multiple classes of modified theories of gravity in the context of cosmology~\cite{Belgacem:2018lbp, Mancarella_2022}. 

GW observations will also probe the large scale structure of the universe. The inhomogeneities in the spatial distribution of the observed compact binary mergers will be another tracer of the large scale structure, which can be measured by the two-point (and higher order) correlation functions~\cite{Vijaykumar_2023}. This will complement the large-scale galaxy and quasar surveys using EM telescopes: GW observations will probe much deeper in redshift, although their spatial resolution will be much poorer than the EM surveys. GW observations of BNSs will be able to accurately measure the scale of baryon acoustic oscillations~\cite{Kumar_2022}, providing an independent probe of the cosmological model. 

In this paper, we discuss a new cosmological probe that GW observations will enable, making use of strongly lensed GWs. GR predicts that intervening massive objects such as galaxies and clusters between the GW source and the observer will deflect the GWs through the phenomenon of gravitational lensing. If the lensing objects are sufficiently massive and compact, and lie sufficiently close to the line of sight to the source, they can create multiple images of the source. This phenomenon, called strong gravitational lensing, is routinely observed in EM observations of galaxies, clusters, quasars, etc. The same objects should strongly lens GW signals as well. The precise fraction of strongly lensed GW sources will depend on the distribution of the GW sources, galaxies and clusters that act as lenses as well as the cosmological model and parameters. According to recent calculations, this fraction is $\sim 0.01-0.05\%$ for current generation detectors~\cite{Li2018,Ng2018} and $\sim 0.1-1\%$ for the XG detectors~\cite{Jana_2023, Xu_2022, Yang_2021, gupta2024characterizing, Oguri_2018,Smith_2023, Wierda_2021, haris2018identifying, wempe2022lensing}. Since XG detectors are expected to detect millions of compact binary mergers during their operation, they will detect thousands or tens of  thousands of strongly lensed GWs. 

Although the theory of lensing of GWs is essentially similar to the lensing of EM waves, there are some important practical differences. Thanks to their short wavelengths, EM observations will allow us to spatially resolve the multiple images of strongly lensed objects. However, multiple images of strongly lensed GW sources cannot be spatially resolved due to their poor sky localisation. In contrast, compact binary mergers produce transient GW signals, which can be temporally localised to milliseconds.  This means that the time delay between the lensed images of the same merger can be measured with exquisite precision using GW observations, which is hard to do in EM lensing. 

The idea of the new method is that the precise number of strongly lensed GW events and the distribution of their lensing time delay depends on the cosmological model and parameters, apart from the distribution of GW sources and lenses~\cite{Jana_2023}.  If the latter two are known, then this will provide a new means of probing the cosmological parameters. The distribution of GW sources can be accurately determined by the large number of un-lensed signals that will dominate the data (see, e.g.,~\cite{Vitale_2019}). The knowledge of the distribution of large gravitational lenses should come from cosmological simulations and EM observations~\footnote{See \cite{Xu_2022} for a complementary approach to constrain the distribution of the GW sources and lenses from the lensing rate and time delay distribution, assuming a cosmology.}. However, we will also show that the cosmological parameters and the model of the distribution of the lenses have sufficiently different imprints on the number of strong lensed events and their time delays, so that it is possible to disentangle these effects to a very good extent. 

Our statistical method, based on the observation of a large number of strongly lensed BBH events, does not require the presence of EM counterparts to the mergers. A complementary approach, presented in~\cite{Liao:2017ioi,LISA_cosmography, Otto_localise_h0}, makes use of a much smaller number of strongly lensed mergers having an EM counterpart. The concurrent observation of lensed EM and GW signals will allow us to measure cosmological parameters from even a single event. However, due to the limited observing horizon of EM telescopes, such observations will be able to probe only low-redshift ($z \lesssim 0.5$) cosmology\footnote{Combing GW lensing observations with lenses detected in EM surveys (e.g., Euclid, CSST and JWST) could probe cosmology at moderate redshifts ($z \lesssim 2$) ~\cite{wempe2022lensing,shan2023microlensing}. However, the exact method for this approach is yet to be developed.}. 

This paper is organised as follows: In section~\ref{sec:lensing_cosmo} we briefly review the Bayesian method that we use to constrain the cosmological parameters from the observation of a population of strongly lensed GW signals. This method was first presented in~\cite{Jana_2023}. In section~\ref{sec:zdist}, we forecast the expected constraints on cosmological parameters assuming various astrophysical models for the redshift distribution of BBH mergers. In the remaining sections we investigate the various sources of systematic errors in our analysis, and show that it is possible to bring them sufficiently small so that interesting measurements are possible in the future. {In particular, in section~\ref{sec:measurement-error}, we explore how the errors in the measurement of the luminosity distance of individual GW events will limit our ability to reconstruct the true redshift distribution of GW sources, hence biasing our cosmological inference.} In section~\ref{sec:lens_dist}, we investigate how our inaccurate understanding of the lens distributions will bias our cosmological inference. In section~\ref{sec:contamination} we will explore how we can deal with contaminated data -- that is, we develop a formalism deal with the presence of a small number of unlensed GW signals in our lensing data, that are misidentified as lensed events due to our limited ability to distinguish between lensed and unlensed GW signals. 

\section{Cosmography using strongly lensed gravitational waves}
\label{sec:lensing_cosmo}

\subsection{Bayesian inference of cosmological parameters}
\label{sec:bayesian_inf}

We assume that $N$ lensed BBH mergers have been detected within an observation period $T_{\mathrm{obs}}$. In this paper, we assume a singular isothermal sphere (SIS) lens model. Thus, each will produce two lensed copies of the GW signal. We also assume that these two images are detected, from which the lensing time delays have been measured accurately. Since the time delays are measured with millisecond precisions we can take them as point estimates, which we denote as $\lbrace \Delta t_i \rbrace_{i=1}^N$. Given $N$ and $\lbrace \Delta t_i \rbrace_{i=1}^N$, we wish to compute the posterior distribution of the cosmological parameters $\vOmega \equiv \{H_0, \Omega_m\}$ assuming a flat $\Lambda$CDM cosmological model.  Using Bayes' theorem:
\begin{equation}
	p\left(\vOmega~|~N,\lbrace \Delta t_i \rbrace, \Tobs, \mathcal{M} \right) = \frac{p \left(\vOmega ~ | ~ \mathcal{M}\right) p\left(N,\lbrace \Delta t_i \rbrace~|~\vOmega,\Tobs, \mathcal{M}  \right)}{p\left(N,\lbrace \Delta t_i \rbrace ~|~\Tobs,  \mathcal{M}\right)}
	\label{eq:posterior}
\end{equation}
where $p\left(\vOmega ~ | ~ \mathcal{M}\right)$ is the prior distribution of 
$\vOmega$ given some model $\mathcal{M}$ while $p\left(N,\lbrace \Delta t_i \rbrace~|~\vOmega, \Tobs,  \mathcal{M}\ \right)$ 
is the likelihood of observing $N$ lensed events with time delays $\lbrace \Delta t_i \rbrace_{i=1}^N$ given the set of cosmological parameters $\vOmega$ and model $\mathcal{M}$.  The normalisation constant ${p\left(N,\lbrace \Delta t_i \rbrace ~|~ \mathcal{M}\right)}$ is the evidence of the assumed model $\mathcal{M}$.
\begin{equation}
	{p\left(N,\lbrace \Delta t_i \rbrace ~|~ T_\mathrm{obs}, \mathcal{M}\right)} = \int p\left(\vOmega ~ | ~ \mathcal{M}\right)p\left(N,\lbrace \Delta t_i \rbrace~|~\vec{\Omega},T_{\mathrm{obs}}, \mathcal{M}  \right) d\vOmega. 
	\label{eq:evidence}
\end{equation}
Above, $\mathcal{M}$ denotes a variety of model assumptions that we employ, including the cosmological model, models of the mass distribution of dark matter halos that act as lenses, lens models, etc. From here onwards, we will drop $\mathcal{M}$ from the expressions, for simplicity of notation. 

Since $N$ and $\lbrace \Delta t_i \rbrace$ are independent data, the likelihood can be written as a product of likelihoods of measuring $N$ lensed events and the set of time delays $\lbrace \Delta t_i \rbrace_{i=1}^N$. 
\begin{equation}
	p\left(N,\lbrace \Delta t_i \rbrace~|~\vec{\Omega},T_{\mathrm{obs}} \right) = p\left(N ~|~\vec{\Omega},T_{\mathrm{obs}}\right) p\left(\lbrace \Delta t_i \rbrace_{i=1}^N ~|~\vec{\Omega},T_{\mathrm{obs}} \right).
\end{equation}
Here, the likelihood of observing $N$ lensed BBH mergers can be described by a Poisson distribution with mean $\Lambda(\vec{\Omega},T_{\mathrm{obs}})$. 
\begin{equation}
	\label{eq:Nlikelihood}
	p\left(N ~|~\vec{\Omega},T_{\mathrm{obs}}\right) = \frac{\Lambda(\vec{\Omega},T_\mathrm{obs})^N\;e^{-\Lambda(\vec{\Omega},T_{\mathrm{obs}})}}{N!}. 
\end{equation}
Above, $\Lambda(\vec{\Omega},T_{\mathrm{obs}})$ is the expected total number of lensed events within the observation period as predicted by the cosmological model with parameters $\vec{\Omega}$. Assuming that  BBH mergers are independent events, the likelihood for observing the set of time delays $\lbrace \Delta t_i \rbrace_{i=1}^N$ can be written as the product of individual likelihoods. 
\begin{equation}
	p\left(\lbrace \Delta t_i \rbrace_{i=1}^N~|~\vec{\Omega},T_{\mathrm{obs}}\right) = 
	\prod_{i=1}^{N} p\left(\Delta t_i~|~\vec{\Omega},T_{\mathrm{obs}}\right). 
\end{equation}
Above, $p(\Delta t_i~|~\vec{\Omega},T_{\mathrm{obs}})$, can be thought of as a ``model'' time-delay distribution $p(\Delta t~|~\vec{\Omega},T_{\mathrm{obs}})$  evaluated at the measured $\Delta t_i$ of a lensed merger. The shape of the model distribution is governed by the cosmological parameters $\vec{\Omega}$. The model distribution $p(\Delta t~|~\vec{\Omega},T_{\mathrm{obs}})$ is obtained from the expected (intrinsic) time delay distribution $p(\Delta t~|~\vec{\Omega})$, after applying the condition that we can not observe the time delays which are greater than the observation time $T_{\mathrm{obs}}$:
\begin{equation}\label{selection-function}
	p\left(\Delta t~|~\vec{\Omega},T_{\mathrm{obs}}\right) \propto p\left(\Delta t~|~\vec{\Omega}\right) {\left(T_{\mathrm{obs}}-\Delta t\right)} \, \Theta( T_\mathrm{obs} - \Delta t),
\label{eq:P_DeltaT_obs_lens}
\end{equation}
where $\Theta$ denotes the Heaviside step function.  

\subsection{Modelling the expected number of lensed events and lensing time delays}
\label{sec:lens_pol_modeling}

The Bayesian inference presented in section~\ref{sec:bayesian_inf} essentially involves comparing the observed number of lensed events $N$ and the distribution of their time delays $\lbrace \Delta t_i \rbrace_{i=1}^N$ with the theoretical prediction of the expected number of lensed events $\Lambda(\vec{\Omega},T_{\mathrm{obs}})$ and their time delay distribution,  $p (\Delta t~|~\vec{\Omega})$, as a function of the parameters $\vec{\Omega}$. Here we describe  how these quantities can be modelled using a cosmological model. We assume the flat $\Lambda$CDM model. However, similar calculations can be performed using more general cosmological models as well.  

\subsubsection{Expected number of lensed events:}

To compute the expected number of lensed binaries, we convolve the redshift distribution of merging binaries with the strong lensing probability at that source redshift. 
\begin{equation}\label{eq:Lambda}
	\Lambda(\vec{\Omega},T_{\mathrm{obs}}) = \mathcal{S}(T_{\mathrm{obs}})  ~ \times ~ R  \int_0^{z_s^\mathrm{max}(\vOmega)} {p_b}(z_s | \vec{\Omega})  \, P_\ell(~z_s|\vec{\Omega}) \, d z_s ,
	\label{eq:exp_num_lens_events}
\end{equation}
Above, $R$ is the BBH detection rate, $p_{b}(z_s | \vec{\Omega})$ is the redshift distribution (probability density) of merging binaries and $P_\ell(z_s | \vec{\Omega})$ is the strong lensing probability for the source redshift $z_s$.  Here we assume that the GW detectors are able to detect all the merging binaries out to $z_{\mathrm{max}}$. For XG detectors, this is a good assumption for the $z_\mathrm{max}$ values that we use~\footnote{The $z_{\mathrm{max}}$ predicted by a source population model (e.g., \cite{Dominik_2013}) assumes the standard cosmology   $\vOmega_\mathrm{true}$. For the population models that we consider, $z_{\mathrm{max}} \simeq 20$. When we consider other values of $\vOmega$, we rescale  $z_\mathrm{max}$ appropriately.}. $\mathcal{S}(T_{\mathrm{obs}})$ denotes the selection effects due to the finite observing time 
\begin{equation}\label{eq:Lambda}
	\mathcal{S}(T_{\mathrm{obs}}) =  \int_{\Delta t=0}^{T_{\mathrm{obs}}} p(\Delta t|\vec{\Omega}) \, (T_{\mathrm{obs}}-\Delta t) \, d\Delta t.
\end{equation}
This takes into account the fact that if the lensing time delay $\Delta t$ is comparable to the observing time $T_\mathrm{obs}$ the second (first) image will be be missed unless the first (second) image arrives at the beginning (end) of the observing run. 

We expect that the rate $R$ of the BBH mergers and their redshift distribution $p_{b}(z_s | \vec{\Omega})$ will be accurately measured from the large ($\sim 10^6$) number of unlensed events that will dominate the data\footnote{What we measure from GW observations is the distribution $p_{b}(d_L)$ of luminosity distance of the sources, which can then be converted into a redshift distribution $p_{b}(z_s | \vec{\Omega})$ assuming a set of cosmological parameters $\vOmega$. For the forecast analysis presented in this paper, we create a luminosity distance distribution from a redshift distribution model, assuming standard cosmological parameters $\vec{\Omega}_\mathrm{true}$. By varying $\vOmega$, we can then obtain different redshift distributions which we use to model the time delay distribution for that specific value of $\vOmega$.}. However, these quantities are currently poorly known. To forecast the expected precision in measuring the cosmological parameters, we take different theoretical models of $p_{b}(z_s | \vec{\Omega})$ assuming a BBH detection rate of $R = 5\times10^5$ per year. The corresponding forecasts are described in section~\ref{sec:zdist}. In section~\ref{sec:measurement-error} we explore the effect of the GW measurement errors in the reconstruction of $p_{b}(z_s | \vec{\Omega})$ and hence on the inference of cosmological parameters. 

To compute the expected number of lensed events using equation~(\ref{eq:exp_num_lens_events}), we also need to know the  probability $P_\ell(~z_s|\vec{\Omega})$ that a source at redshift $z_s$ is strongly lensed. This will depend on the distribution of lenses as well as cosmological parameters. We assume that the lenses are modelled by the SIS model. Multiple images are produced when the projected location of the source in the lens plane is within the Einstein radius of the lens, given by 
\begin{equation}
	r_E(\sigma, \zl, \zs, \vOmega) = 4 \pi \left(\frac{\sigma}{c}\right)^2 \frac{\Dl(\zl, \vOmega) ~ \Dls(\zl, \zs, \vOmega)}{\Ds(\zs, \vOmega)},
\end{equation}
where $\sigma$ is the velocity dispersion of the lens, while $\Dl, \Ds$ and $\Dls$ are the angular diameter distance to the lens, to the source, and between the lens and the source, respectively. The strong lensing probability is given by integrating the differential optical depth for strong lensing by different lenses  
\begin{equation}
	P_\ell(z_s, \vOmega) = \int_0^{z_s} \int_{\sigma_\mathrm{min}}^{\sigma^\mathrm{max}} \frac{d\tau}{dz_\ell d\sigma} (\zs, \zl, \sigma, \vOmega) ~ dz_\ell d\sigma, 
	\label{eq:opt_depth}
\end{equation}
where the differential optical depth for strong lensing by a lens with velocity dispersion $\sigma$ located at a redshift $\zl$ is given by the fraction of the full sky covered by lenses 
\begin{equation}
	\frac{d\tau}{dz_\ell d\sigma} (\zs, \zl, \sigma, \vOmega) = \frac{dV_c}{d\zl}(\zl, \vOmega) ~\times~ \frac{dN_\ell}{dV_c d\sigma}(\zl, \sigma, \vOmega) ~\times~ \frac{\pi r_E^2(\sigma, \zl, \zs, \vOmega)}{4 \pi \Dl^2(\zl, \vOmega)}, 
	\label{eq:diff_opt_depth}
\end{equation}
where 
\begin{equation}
	\frac{dV_c}{d\zl}(\zl, \vOmega)  = \frac{4 \pi c}{H_0} \frac{(1+\zl)^2 \, \Dl^2(\zl, \vOmega)}{E(\zl, \vOmega)}
	\label{eq:diff_cvol}
\end{equation}
is the differential comoving volume and 
\begin{equation}
	\frac{dN_\ell}{dV_c d\sigma} (\zl, \sigma, \vOmega) = n_c(\zl, , \vOmega) ~ p_\sigma(\sigma, \zl)
	\label{eq:lens_dist}
\end{equation}
is the comoving number density of lenses with velocity dispersion $\sigma$ at redshift $\zl$. Here, $p_\sigma(\sigma, \zl)$ is the probability density of the dispersion velocity of the lenses at a redshift $\zl$ and  $n_c(\zl)$ is the comoving number density of lenses at $\zl$. In equation~(\ref{eq:diff_cvol}), $E(\zl, \vOmega) = \sqrt{\Omega_M[(1+\zl)^3-1]+1}$ assuming a flat $\Lambda$CDM cosmology. 

In order to compute the distribution of lenses at a given redshift, we need to use some models of structure formation.  
\begin{equation}
	\frac{dN_\ell}{dV_c d\sigma} (\zl, \sigma, \vOmega) = \frac{dN_\ell}{dV_c dM} (\zl, M, \vOmega) ~\times~ \frac{dM}{d\sigma}, 
\end{equation}
where the first term denotes the comoving number density of dark matter halos with mass $M$ at redshift $\zl$ predicted by the cosmological model with parameters $\vOmega$. We consider several models of the halo mass function (HMF) calibrated to cosmological simulations. The second term is a Jacobian to convert the distribution of the halo mass to that of the dispersion velocity. To compute the Jacobian, we assume that the halos are spherically symmetric and virialised, with uniform density $\rho$ and radius $R$. Thus,  
\begin{equation}
	\sigma \simeq \sqrt{\frac{GM}{R}}, ~~~ M = \frac{4}{3}\pi R^3 \rho ~~~ \Rightarrow \frac{dM}{d\sigma} = \frac{3M}{\sigma}. 
\end{equation}
We also need to use some minimum and maximum cutoff for $\sigma$ to compute the total optical depth defined in equation~(\ref{eq:opt_depth}). The natural choices are $\sigma_\mathrm{min} = \sigma(M_\mathrm{min})$ and $\sigma_\mathrm{max} = \sigma(M_\mathrm{max})$. We assume $M_{\mathrm{min}}=10^8\;M_{\odot}$ and $M_{\mathrm{max}}\simeq10^{15}\;M_{\odot}$, since this the mass range of validity of most of the HMF models that we use. 

It is now easy to see from equation~(\ref{eq:diff_opt_depth}) why the cosmological parameters affect the lensing optical depth and hence the number of detected lensed events. The first term describes a purely geometrical effect of how the comoving volume at a given redshift varies with a change in cosmological parameters. The second term describes the change in the distribution of lenses due to changes in the structure formation. Third shows how the fractional area covered by lenses at a given redshift varies due to the geometric effect.  

\subsubsection{Expected distribution of lensing time delays:}

In the SIS lens model, the time delay between the two images is given by (see, e.g. \cite{Oguri_2002}):
\begin{equation}\label{delta-t-sis}
	\Delta t (\zl, \sigma, \zs, y, \vec{\Omega})  =  \frac{32 \, \pi^2 \, y}{c} ~ \left(\frac{\sigma}{c}\right)^4 ~ (1+\zl) ~\frac{\Dl(\zl, \vOmega) \Dls(\zl, \zs, \vOmega)}{\Ds(\zs, \vOmega)}, 
\end{equation}
where $y$ is the projected location of the source on the lens plane (in units of $r_E$). We compute the expected time delay distribution $p (\Delta t~|~\vec{\Omega} )$ for different values of the cosmological parameters $\vec{\Omega}$ by marginalising the distribution of time delay over all other parameters $\vec{\lambda} \equiv \{y,\sigma,z_{\ell},z_s\}$ on which the time delay depends. 
\begin{equation}\label{delta-t-integration}
	p\left(\Delta t ~|~ \vec{\Omega}\right) = \int p\left(\Delta t ~|~ \vec{\lambda},\vec{\Omega}\right) p (\vec{\lambda}~|~\vec{\Omega}) \, d\vec{\lambda},
\end{equation}
where $p(\vec{\lambda}~|~\vec{\Omega})$ denotes the expected distribution of the source position $y$, lens velocity dispersion $\sigma$, lens redshift $z_{\ell}$ and source redshift $z_s$, given the set of cosmological parameters $\vec{\Omega}$. If we assume isotropy of space, the distribution of $y$ is independent of the cosmological parameters. Hence 
\begin{eqnarray}
	p(\vec{\lambda}~|~\vec{\Omega}) = p(y) ~ p(\sigma, \zl, \zs ~ | ~ \vOmega), 
\end{eqnarray}
where $p(y) \propto y$ with $y = (0, 1]$. Above, $p(\zl, \sigma, \zs ~ | ~ \vOmega)$ can be further split as 
\begin{equation}
	p(\sigma, \zl, \zs ~ | ~ \vOmega) = p(\sigma, \zl ~ | ~ \zs , \Omega)  ~ p_b(\zs ~ | ~ \vOmega), 
	\label{eq:P_sigma_zl_zs_given_Omega}
\end{equation}
where $p_b(\zs ~ | ~ \vOmega)$ is the expected/measured distribution of source redshifts, while $p(\sigma, \zl ~ | ~ \zs , \Omega)$ is computed from the differential optical depth  [equation~(\ref{eq:diff_opt_depth})]
\begin{equation}
	p(\sigma, \zl ~ | ~ \zs , \vOmega) \propto \frac{d\tau}{d\zl d\sigma} (\zs, \vOmega). 
	\label{eq:P_sigma_zl}
\end{equation}
Thus, the essential ingredients for modelling the expected number of lensed events and their time delay distribution are: 
\begin{itemize}
	\item \emph{The redshift distribution of GW sources:} We expect that this can be measured with sufficient precision from the large number $\sim 10^6$ of unlensed events that will dominate the data (see, e.g.,~\cite{Vitale_2019}). In section~\ref{sec:zdist} we forecast the prospective constraints on cosmological parameters assuming various theoretical models of the source redshift distribution (figure~\ref{fig:zdist}). {In section~\ref{sec:measurement-error} we study how uncertainties and errors in inferring the source redshift distribution can affect the constraints on cosmological parameters (figure \ref{fig:measurement-error-posterior}).} 
	\item \emph{A halo mass function model:} This will need input from cosmological simulations. We show in section~\ref{sec:lens_dist} that a wrong choice of the HMF model can bias our inference of cosmological parameters. However, if the right HMF model is one among the many models that we consider, Bayesian model selection can be used to identify the right model. 
\end{itemize}

\section{Effect of source distribution models}
\label{sec:zdist}

The redshift distribution of sources is a significant input for our method. The precision of estimation of cosmological parameters depends on the redshift distribution of sources because lensing optical depth increases with redshift [equation~(\ref{eq:opt_depth})]. To examine how the source population affects the precision in estimating the cosmological parameters, we conduct a recovery test similar to the one done in \cite{Jana_2023}. We examine various models for the redshift distribution of mergers, including those predicted by population synthesis models such as \cite{Belczynski_2016a,Dominik_2013,Belczynski_2016nat}, as well as a model in which the merger rate is uniform in comoving volume. Additionally, we explore merger distribution models obtained from a star formation rate given in \cite{Madau_Dickinson_2014} using different delay time distributions as presented in \cite{Vitale_2019}. We consider two different models for the distribution of time delays between the formation of the binary and its merger: an exponential distribution with a characteristic time scale $\tau = 0.1\;\mathrm{Gyr}$ and a distribution uniform in logarithmic of time delay (see equations $6$ and $7$ of \cite{Vitale_2019}). Apart from these two models, we also consider the scenario where there is no time delay between formation and merger, which implies that black hole mergers follow the same redshift distribution as the star formation rate. 

\begin{figure}[tbh]
	\includegraphics[width=0.325\columnwidth]{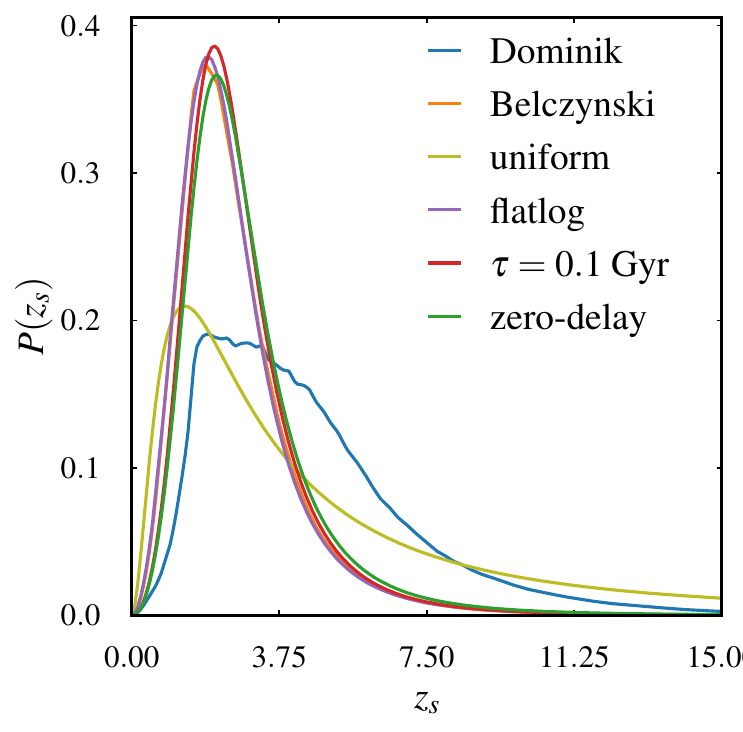}
	\includegraphics[width=0.342\columnwidth]{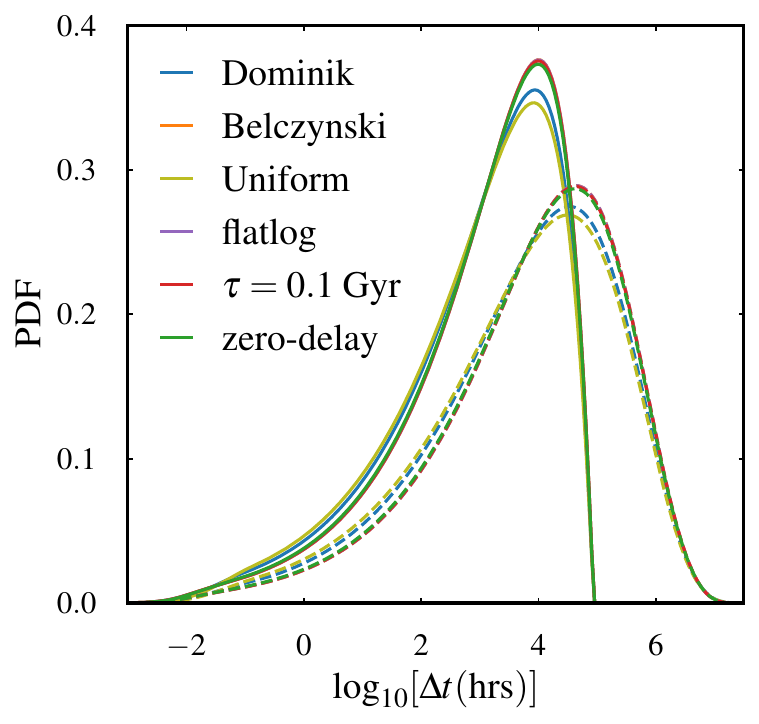}
	\includegraphics[width=0.322\columnwidth]{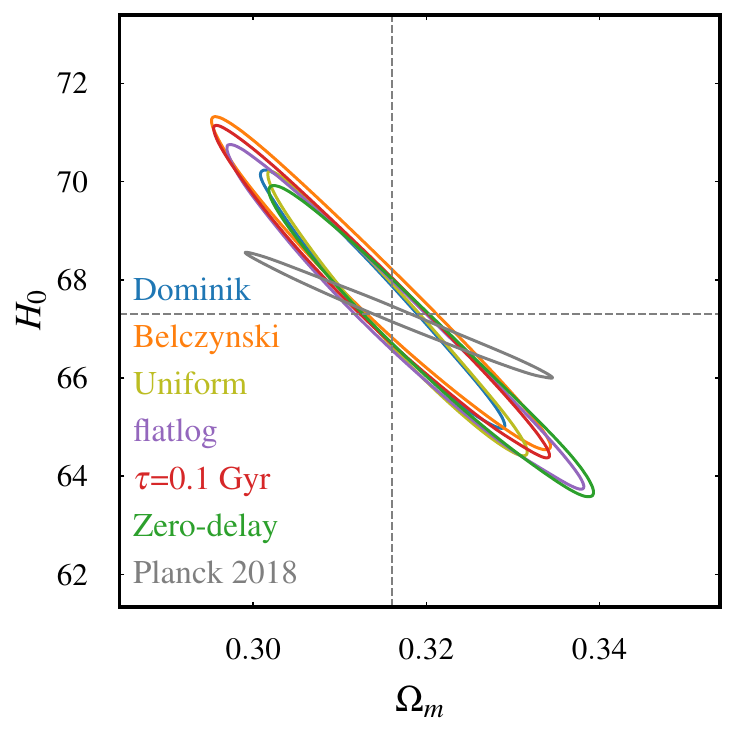}
	\caption{\emph{Left:} Different models for the distributions of redshifts for BBH mergers. These include prediction from population synthesis studies by Dominik~\cite{Dominik_2013} and Belczynski~\cite{Belczynski_2016a,Belczynski_2016nat}, as well as uniform in comoving volume. Other models are based on Madau-Dickinson star formation rate~\cite{Madau_Dickinson_2014} and consider different distributions of the time delay between formation and merger. \emph{Middle:} The distributions of time delay between two lensed images of a source for different models of redshift distribution of BBH mergers shown in the left panel. Dashed lines represent the actual time delay distributions $p(\Delta t~|~\vec{\Omega})$ while solid lines represent time delay distributions $p(\Delta t~|~\vec{\Omega},T_{\mathrm{obs}})$ that will be observable in $10\;\mathrm{yrs}$. Time delay distributions are calculated for $\vec{\Omega} = \{\Omega_m=0.316,\;H_0=67.3\}$. \emph{Right:} Posteriors (95\% credible region) on cosmological parameters $\left(\Omega_m, H_0\right)$ for different models of redshift distributions. This is done considering a BBH merger rate of $5\times10^5\;\mathrm{yr}^{-1}$ with observation time period of $10\;\mathrm{yrs}$. Dashed cross represents true cosmological parameters.}
	\label{fig:zdist}
\end{figure}

The left panel of figure \ref{fig:zdist} displays the redshift distributions of BBH mergers which we consider. We consider a ``true'' cosmology $\vec{\Omega}_{\mathrm{true}} = \{\Omega_m=0.316,\;H_0=67.3\}$. The true distributions of lens redshift and parameters are calculated using a HMF  model as described by \cite{Behroozi_2013}, implemented in \textsc{HMFcalc} package \cite{murray2013hmfcalc} as described in section~\ref{sec:lens_pol_modeling}. We assume a merger rate of $R = 5\times10^5\;\mathrm{yr}^{-1}$ for BBHs, with an observation time period $T_{\mathrm{obs}}=10\;\mathrm{yrs}$. 

To simulate one observational scenario where $N$ lensed events with time delays $\left\{\Delta t_i\right\}_{i=1}^{i=N}$  are detected, we draw one sample from a Poisson distribution with mean $\Lambda\left(\vec{\Omega}_{\mathrm{true}},T_{\mathrm{obs}}=10\;\mathrm{yr}\right)$ and then draw samples $\left\{\Delta t_i\right\}_{i=1}^{i=N}$  from $p\left(\Delta t ~|~\vec{\Omega}_{\mathrm{true}},T_{\mathrm{obs}}=10\;\mathrm{yrs}\right)$. We neglect the selection effects of XG detectors, as XG detectors are expected to detect all the BBHs out to very high redshifts. Time delay distributions considering different models of redshift distribution for BBH mergers are shown in the middle panel of figure \ref{fig:zdist} (for $\vec{\Omega} = \vec{\Omega}_{\mathrm{true}}$). 

\begin{table}[h]
	\centering
	\begin{tabular}{ c c c c c c c } 
		\hline
		\hline 
		& Dominik & Belczynski & Uniform& flatlog &$\tau = 0.1\mathrm{Gyr}$& zero-delay \\ 
		\hline
		\hline 
		$\Lambda\left(\Omega_{\mathrm{true}}\right)$& 37700 & 20594 & 36698& 20448 &22098&23433\\ 
		$\Omega_m(68\%)$& $0.315^{+0.006}_{-0.006}$ & $0.314^{+0.007}_{-0.008}$ &$0.316^{+0.006}_{-0.006}$ &$0.317^{+0.008}_{-0.009}$   &$0.314^{+0.008}_{-0.008}$&$0.320^{+0.007}_{-0.008}$\\ 
		$\Omega_m(95\%)$& $0.315^{+0.012}_{-0.011}$ & $0.314^{+0.016}_{-0.015}$ &$0.316^{+0.012}_{-0.012}$ &$0.317^{+0.017}_{-0.016}$   &$0.314^{+0.016}_{-0.015}$&$0.320^{+0.016}_{-0.014}$\\ 
		$H_0(68\%)$& $67.6^{+1.1}_{-1.1}$ & $67.9^{+1.4}_{-1.4}$ &$67.2^{+1.2}_{-1.2}$ & $67.2^{+1.4}_{-1.4}$ &$67.7^{+1.4}_{-1.4}$&$66.7^{+1.3}_{-1.3}$\\ 
		$H_0(95\%)$& $67.6^{+2.1}_{-2.0}$ & $67.9^{+2.7}_{-2.7}$ &$67.2^{+2.4}_{-2.3}$ & $67.2^{+2.9}_{-2.7}$ &$67.7^{+2.7}_{-2.6}$&$66.7^{+2.5}_{-2.5}$\\ 
		\hline
		\hline
	\end{tabular}
	\caption{Expected number of lensed events $\Lambda$ for $\vec{\Omega}=\vec{\Omega}_{\mathrm{true}}$ and expected constraints (68\% and 95\% credible intervals) on $\Omega_m$ and $H_0$ for different models of redshift distribution of BBH mergers considering a merger rate of $5\times10^5\;\mathrm{yr}^{-1}$ with an observation time of $10\;\mathrm{yrs}$.}
	\label{tab:zdist}
\end{table}

Using the method outlined in section~\ref{sec:bayesian_inf}, we compute the posteriors on the cosmological parameters $\vOmega$ from the different observing scenarios corresponding to the different source redshift distributions. We also assume that the redshift distribution of the sources is measured with sufficient precision from the observation of unlensed events. Posteriors for the cosmological parameters are shown in the right panel of figure \ref{fig:zdist}. When the source populations extend to high redshift (e.g., Dominik, Uniform), the precision is better in comparison to populations where the merger redshift does not have support at high redshift (e.g., Belczynski, flatlog, etc.). This is because the lensing optical depth increases with redshift, and therefore, populations that extend to high redshift are expected to have a larger number of lensed events than the populations that don't extend to high redshift. The expected 95\% credible intervals in the posteriors of the cosmological parameters are summarised in table \ref{tab:zdist}, for different source redshift distribution models. 

\section{Effect of errors in measuring the source redshift distribution}
\label{sec:measurement-error}

\begin{figure}[tbh]
	\includegraphics[width=0.33\columnwidth]{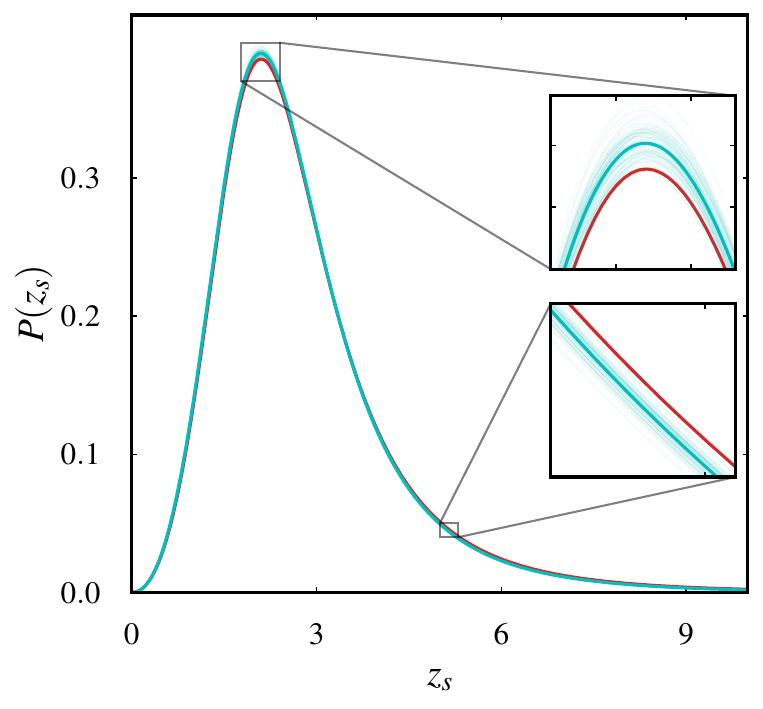}
	\includegraphics[width=0.335\textwidth]{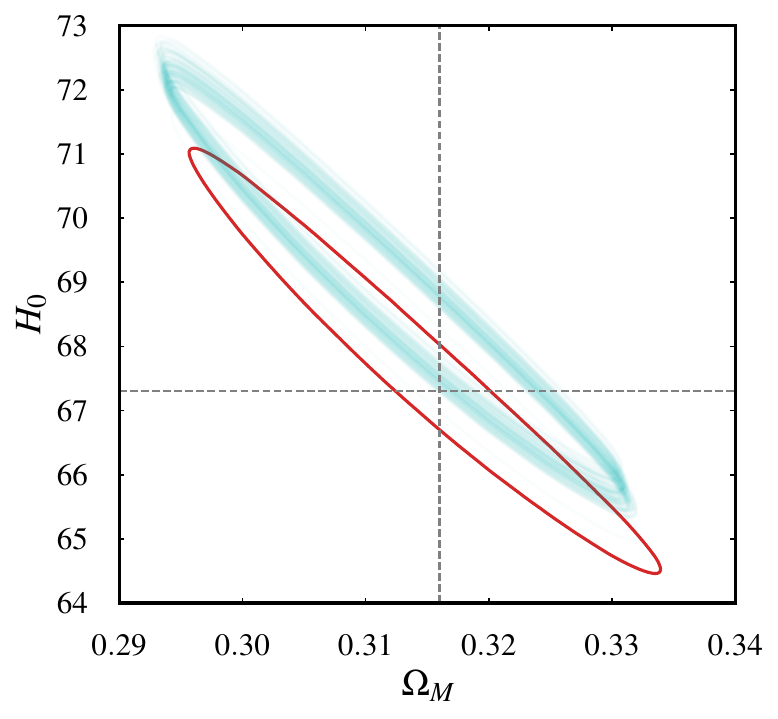}
	\includegraphics[width=0.335\textwidth]{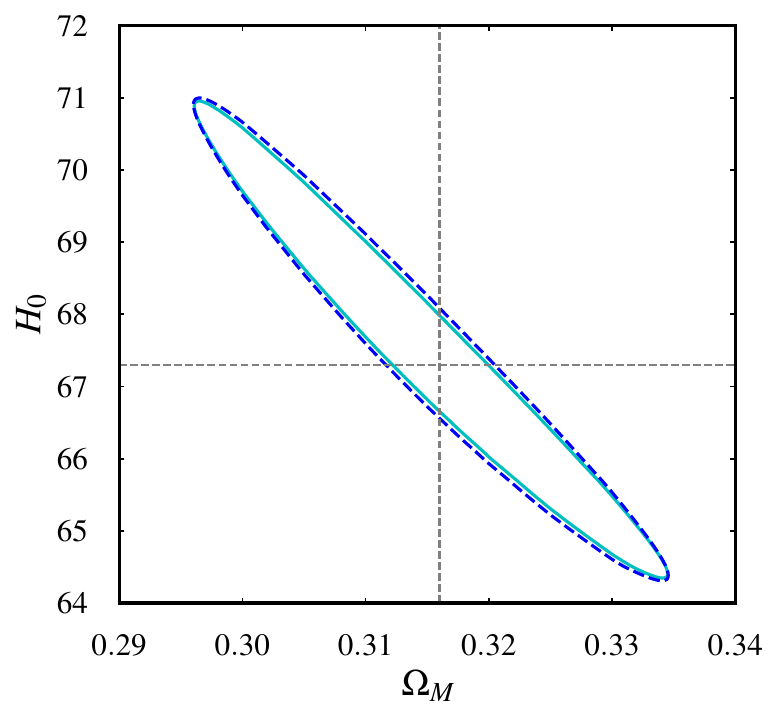}
	
	\caption{\emph{Left panel:} The red curve shows the redshift distribution model based on the Madau-Dickinson star formation rate, with an exponential delay-time distribution with $\tau=0.1\;\mathrm{Gyr}$. Light cyan curves show the reconstructed (posterior predictive) distributions as shown in figure 5 of \cite{Vitale_2019}, while the solid cyan curve shows their average. 
	\emph{Middle Panel:} The 95\% credible regions of the posteriors of $\Omega_m$ and $H_0$. The lensed events are simulated using the red curve in the left panel. The cosmological inference is done with the `true' redshift distribution (red contour), as well as with 100 reconstructed source redshift distributions (light cyan contours, corresponding to the light cyan curves in the left panel). The cyan posteriors are biased due to the biased reconstruction of the source redshift distribution. 
	\emph{Right Panel:} This panel shows the effect of the marginalisation over the source redshift distribution uncertainties. Here, the lensed events are simulated using the solid cyan curve in the left panel, so that there is no systematic bias. The solid cyan contour shows the posterior estimated using the same redshift distribution. The dashed blue contour shows the posterior that is marginalised over the source redshift distribution uncertainties. The broadening of the posterior due to this effect is minimal.}
\label{fig:measurement-error-posterior}
\end{figure}

In section~\ref{sec:zdist}, we assumed that the true redshift distribution of the sources is accurately known from the observation of unlensed events. However, this is not an entirely valid assumption as uncertainties and errors in the measurement of luminosity distance could bias our estimation of the redshift distribution, leading to systematic errors as well as additional statistical uncertainties in the estimation of cosmological parameters. Here we investigate the severeness of this effect. 

Combining the inferred luminosity distance posteriors from a number of BBH events and assuming a cosmology, we will be able to reconstruct the redshift distribution of BBH mergers in the universe. This can be done either using a parametric model of the population properties of BBHs (such as their mass and redshift distribution) or using non-parametric methods~\cite{Rinaldi:2021bhm}. The errors in the reconstruction of the redshift distribution of BBH mergers using XG detectors were studied in~\cite{Vitale_2019}, using a parametric model. We use the results of their study for characterising the corresponding errors in cosmological inference. As done in~\cite{Vitale_2019}, we use a `true' redshift distribution given by a model based on Madau-Dickinson star formation rate and exponential delay time distribution with a time scale, $\tau = 0.1\; \mathrm{Gyr}$ (red curves in figures~\ref{fig:zdist} and \ref{fig:measurement-error-posterior}). 

We use the posterior predictive distributions of the merger rate density from~ \cite{Vitale_2019}, which was derived from the posteriors on the parameters of the redshift distribution model (figures 4 and 5 of \cite{Vitale_2019}). We take $\sim 100$ samples of the merger rate density distributions and convert them into source redshift distributions. In the left panel of figure \ref{fig:measurement-error-posterior}, the light cyan curves represent these posterior predictive distributions while the solid cyan curve shows their average. These reflect the uncertainties in the measurement of the redshift distribution, while the red curve shows the `true' distribution. 

Now we simulate lensed events using the `true' redshift distribution and infer cosmological parameters using different posterior predictive redshift distributions (corresponding to different light cyan lines in the left panel of figure \ref{fig:measurement-error-posterior}). The middle panel of figure \ref{fig:measurement-error-posterior} shows the posteriors of the cosmological parameters inferred using the `true' redshift distribution (red line) and 100 sampled redshift distributions (thin cyan lines). It is evident that there is a systematic bias in the recovery, which is attributed to the biased reconstruction of the merger rate density as seen in the left panel of figure \ref{fig:measurement-error-posterior}. 

We also explore a scenario in which the redshift distribution of BBHs is estimated without any systematic biases, but with some statistical uncertainty. In order to simulate such a scenario, we use the average of 100 sampled redshift distributions as the `true'  distribution (cyan curve in the left panel of figure~\ref{fig:measurement-error-posterior}). We then use this `true' distribution to simulate lensed events and to infer cosmological parameters (right panel of figure \ref{fig:measurement-error-posterior}). Here, the solid cyan contour represents the posterior (95\% credible region) of cosmological parameters using the new `true' redshift distribution, showing no systematic bias. When we factor in the uncertainties in the estimation of source redshift distribution making use of the posterior predictive distributions (light cyan curves in the left panel of ~\ref{fig:measurement-error-posterior}), the posteriors of the cosmological parameters will have a scatter similar to the light cyan curves in the middle panel of figure~\ref{fig:measurement-error-posterior}. The dashed dark blue contour shows the posterior that is marginalised over the uncertainties in the estimation of the source redshift distribution. We can see that the broadening of the posterior is minimal. 

In summary, the expected statistical uncertainties in the estimation of the source redshift distribution have a negligible effect on cosmological inference. However, the redshift distribution needs to be estimated without any systematic bias. In this preliminary investigation, we have neglected the correlation of the parameters of the source distribution model with the cosmological parameters. We anticipate that this oversight will not significantly broaden the posteriors on cosmological parameters, as the expected measurement error on the merger rate density is small~\cite{Vitale_2019}. Our future plans include conducting a comprehensive analysis that considers all the parameters of the source redshift distribution along with the cosmological parameters and marginalising over them. 

\section{Effect of lens distribution}
\label{sec:lens_dist}

\begin{figure}[tbh]
	\includegraphics[width=0.45\columnwidth]{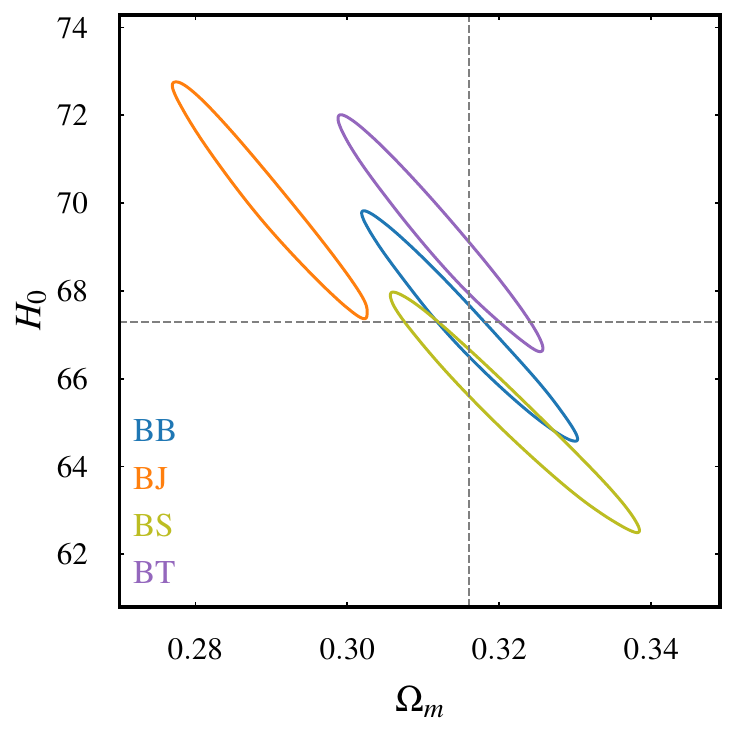}
	\includegraphics[width=0.45\columnwidth]{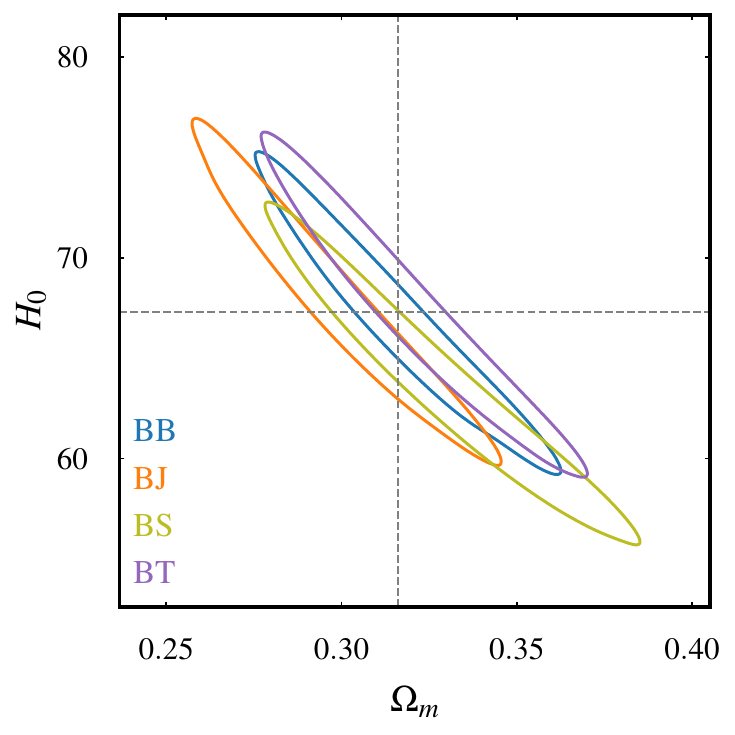}
	\caption{Posteriors (95\% credible regions) of cosmological parameters that illustrate the bias in the inference due to the use of wrong models of the HMF. In the legends, BB,  BJ, BS and BT denote different scenarios when our simulated observation of lensed signals is produced using Behroozi model, and then cosmological parameters are recovered using Behroozi, Jenkins, SMT and Tinker08 models,  respectively. \emph{Left panel:} Results corresponding to a merger rate $5\times10^5\;\mathrm{yr}^{-1}$. We can see that true cosmology is not recovered within 95\% credible region except for the case of BB. However, the precision for all cases remains almost the same. \emph{Right panel:} Results corresponding to a merger rate $5\times10^4\;\mathrm{yr}^{-1}$. Here, we can see that the amount of bias is relatively small due to the decreased precision of the cosmological parameters.}
	\label{fig:lens-dist-bias}
\end{figure}

The distribution of the redshift and other parameters of the lenses are determined using the HMF models (see section~\ref{sec:lens_pol_modeling}). Here we investigate how selecting the incorrect model for the lens distribution (both redshift and other parameters) can affect the inference of cosmological parameters. We broadly follow the same steps as outlined in section \ref{sec:zdist}. However, here we assume the redshift distribution given by Dominik \cite{Dominik_2013} to be the one presented by nature, and we will vary the lens distribution while keeping the other parameters fixed. We consider four different models of HMF, namely Behroozi~\cite{Behroozi_2013}, Jenkins~\cite{Jenkins_2001}, SMT~\cite{SMT_2001}, and Tinker08~\cite{Tinker_2008}. We simulate $N$ lensed events with time delays $\{\Delta t_i\}_{i=1}^N$ using the Behroozi model. $N$ is drawn from a Poisson distribution with mean $\Lambda(\vOmega_\mathrm{true}, T_{\mathrm{obs}})$, where $\vOmega_\mathrm{true}$ is the assumed ``true'' value of cosmological parameters. This is  our simulated observational data. We then estimate the cosmological parameters using all of these HMF models. 

We observe that there is a bias in the inference of cosmological parameters when we choose the wrong model of lens distribution. Figure~\ref{fig:lens-dist-bias} shows an example set of posteriors (95\% credible regions) on the cosmological parameters when different HMF models are used in the parameter estimation.  The left (right) panel shows results corresponding to an assumed merger rate $R = 5\times10^5\;\mathrm{yr}^{-1}$ ($5\times10^4\;\mathrm{yr}^{-1}$) for an observation time of $10\;\mathrm{yrs}$. In the left panel the true cosmology is not recovered within 95\% credible region except when the true HMF is used in the parameter estimation, while the amount of bias is relatively small in the right panel due to the decreased precision. 

\begin{figure}[tbh]
	\includegraphics[width=0.45\columnwidth]{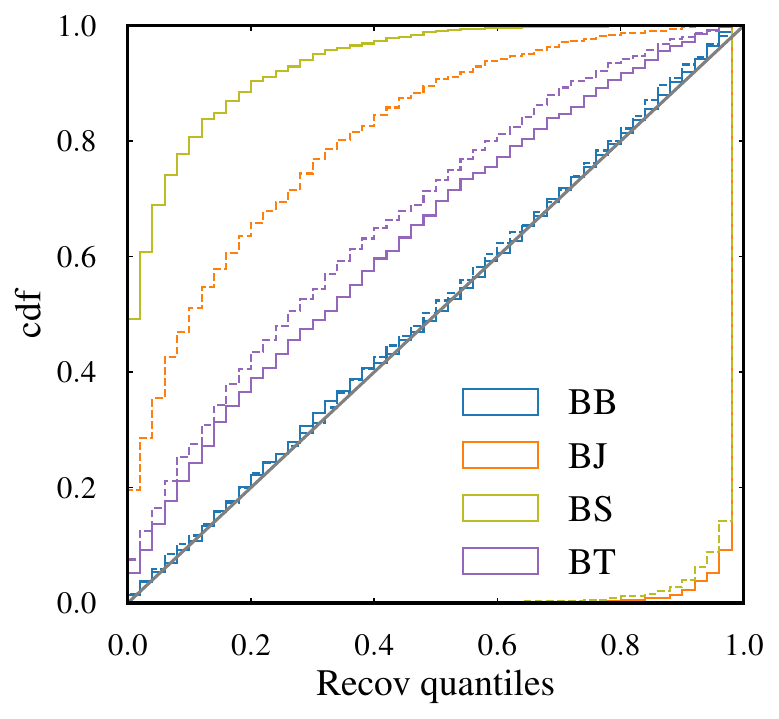}
	\includegraphics[width=0.45\columnwidth]{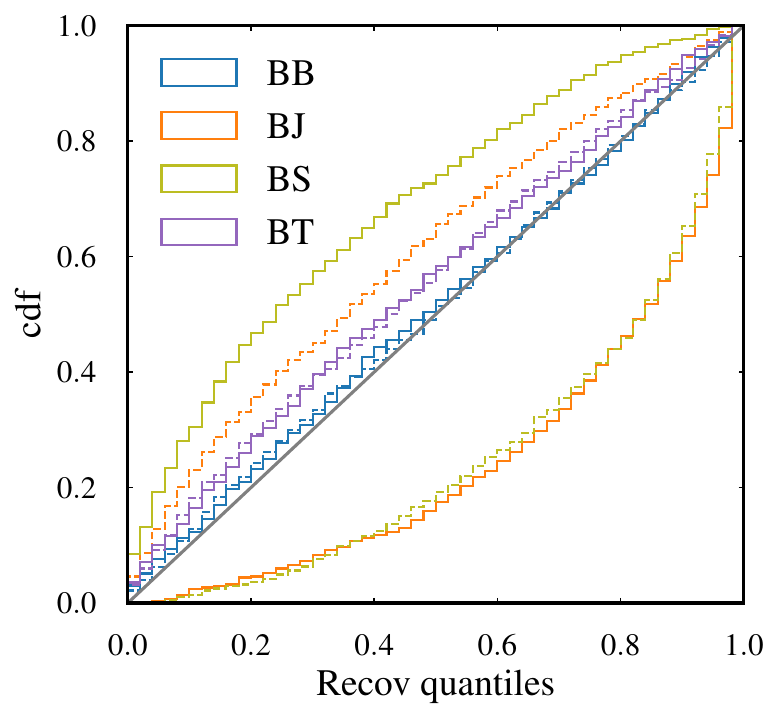}
	\caption{The cumulative distribution of the quantiles (p-p plots) in which the true cosmological parameters are recovered from $\sim 10^3$ recovery tests. The legend box shows the scenarios considered (same as figure~\ref{fig:lens-dist-bias}). Any deviation from the diagonal line suggest there is a bias in the recovery. The larger the deviation, the greater is the bias in the inference. Solid (dashed) lines show the recovery for $\Omega_m$ ($H_0$). In the {left panel}, the results corresponding to a merger rate of $5\times10^5\;\mathrm{yr}^{-1}$ show that there are biases, except in the BB scenario. In the {right panel}, the results for a merger rate of $5\times10^4\;\mathrm{yr}^{-1}$ indicate that there is relatively lower bias due to the decreased precision in the estimation of the cosmological parameters.}
	\label{fig:lens-dist-bias-in-recovery}
\end{figure}

The posteriors in figure~\ref{fig:lens-dist-bias} show one realisation of the observation scenario. Each of the posterior could be randomly shifted due to the finite number of observed events (Poisson fluctuations). To get the statistical nature of the biases, we perform this analysis using $\sim 10^3$ catalogs of observations of same statistical nature.  The so-called probability-probability (p-p) plots show in what fraction of the simulated observations, the true values are recovered within a given credible interval (figure \ref{fig:lens-dist-bias-in-recovery}). If the theoretical models agree with the data, the p-p plot should show a diagonal line.  We see that diagonal p-p plots are obtained only when the simulated HMF model is used in the parameter estimation. 

\begin{figure}[tbh]
	\includegraphics[width=0.45\columnwidth]{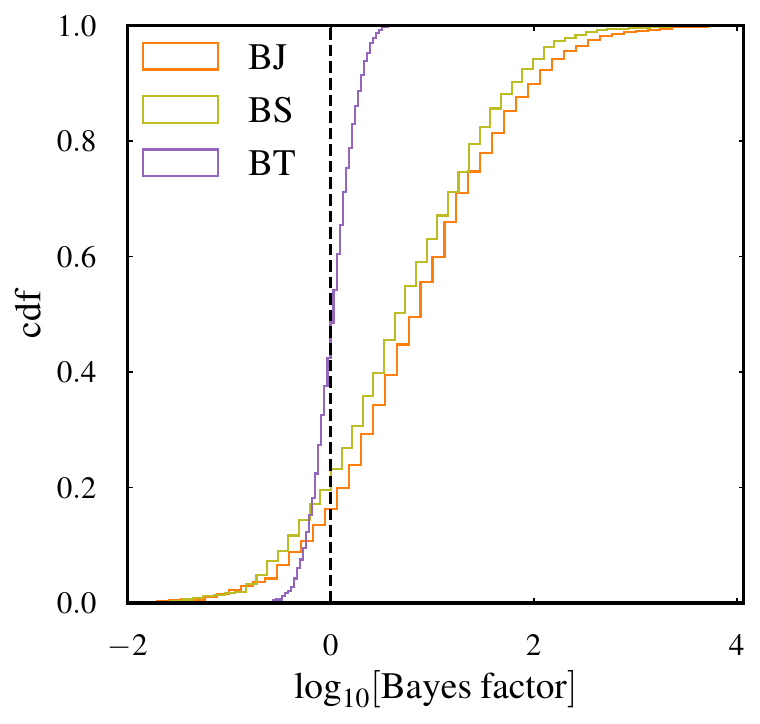}
	\includegraphics[width=0.45\columnwidth]{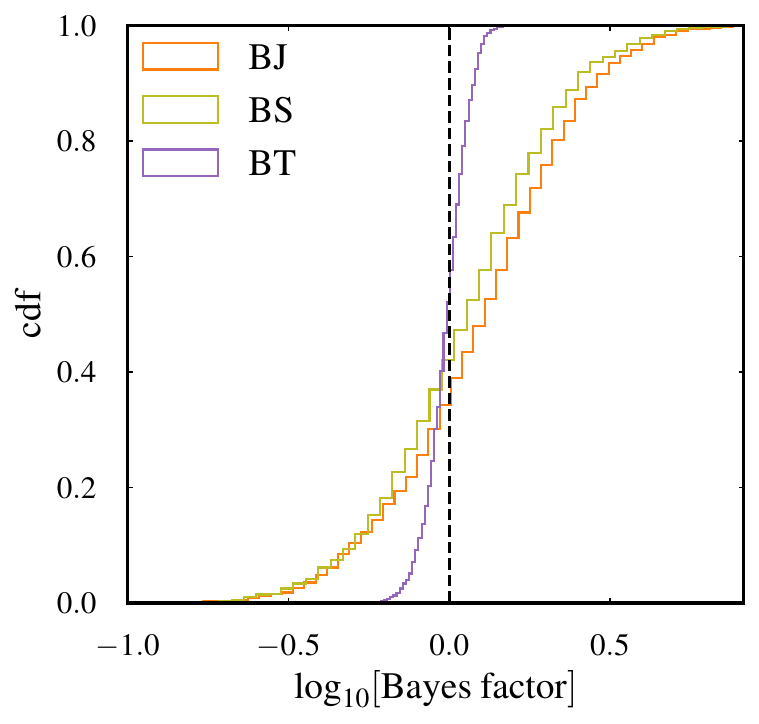}
	\caption{The cumulative distribution of  $\log_{10}$~(Bayes factor) between the ``right'' (B) and ``wrong'' (J, S or T) HMF models, computed from $\sim 10^3$ simulated catalogs. The results correspond to merger rates of  $5\times10^5\;\mathrm{yr}^{-1}$ (left) and $5\times10^4\;\mathrm{yr}^{-1}$ (right). We can see that HMF models that cause significant bias in the parameter estimation have low Bayes factors most of the time.}
	\label{fig:lens-dist-bayes-factor}
\end{figure}

While this is a cause of concern, we show below that Bayesian model selection generally allows us to identify the correct model of the HMF. We compute the ratios of Bayesian evidences [equation~(\ref{eq:evidence})] of different HMF models and show that the right HMF model is almost always preferred. We simulate lensed events with time delays using model B (Behroozi) and then recover the cosmological parameters using all these models. We observe that for the model that consistently shows bias in its parameter estimates, the evidence is smaller than the ``true'' model. The ratio of the evidences (the Bayes factor) between the true and false models is greater than one for most of the simulations. For instance, in the parameter estimation using the J (Jenkins) model typically causes large bias in the estimated parameters (see figure~\ref{fig:lens-dist-bias}) . However, the Bayes factor between B and J models is higher than one over $\sim 80\%$ of the time, indicating that the true model is generally preferred (figure~\ref{fig:lens-dist-bayes-factor}). In the parameter estimation using the T (Tinker) model, the true model is preferred only for $\sim 50\%$ of the time; however, the bias in the estimated parameters using the T model is generally smaller. 

In order for this model selection to work, the space of models that we consider should include the true model also. A more powerful method would be to create parametrised models of the HMF that also incorporate modelling errors as extra parameters. These could be treated as nuisance parameters in the cosmological parameter estimation and marginalised over. A combination of EM observations and cosmological simulations can be used to further improve our models of the distribution of lens properties. 

There are similarity transformation degeneracies that prevent us from reconstructing the lens mass and redshift simultaneously from a single event~\cite{Gorenstein_1988,Saha_2000,Falco_1985, Kochanek_2002}. This is why the GW observations of individual lensed events, in the absence of EM observations, will not enable any measurement of cosmological parameters~\cite{Poon:2024}. However, this doesn’t affect our method in a serious way, since we assume the knowledge of the distribution of the lens properties in the form of a HMF. Even if we don’t know the ``true'' HMF, if the true HMF is in the list of models that we use, it will have the highest Bayes factor statistically. The reason is that the change in the time delay distribution due to a different HMF model cannot be exactly mimicked by a change in cosmology, i.e., the degeneracy between cosmology and HMF is not an exact one.

\section{Effect of contamination}
\label{sec:contamination}

Any data analysis method that is used to identify strongly lensed signals in GW data will have a true positive probability $\epsilon$ and a false positive probability $\alpha$ (see, e.g.,~\cite{haris2018identifying, Janquart_2023, Caliskan_lensing_luck_2023}). This means that some unlensed GW signals will be incorrectly identified as lensed signals and some lensed signals will be missed, thus biasing the detected number of lensed events and their time delay distribution. In this section we develop a strategy that can incorporate this effect, thus evading systematic biases in the estimation of cosmological parameters.  

We assume that the total number $N_\mathrm{tot}$ of detected BBH mergers observed follows a Poisson distribution of mean $\Lambda_\mathrm{tot}$ while the detected number of  strongly lensed mergers $N$ follows a Poisson distribution of mean $\Lambda$. We define the contamination fraction $\kappa$ as the ratio between the expected number of falsely identified lensed pairs and the number of truly identified lensed pairs 
\begin{equation}
	\kappa \simeq \frac{\alpha \Lambda_\mathrm{tot}^2}{2} ~ \frac{1} {\epsilon u \, \Lambda_{\mathrm{tot}}} = \frac{\alpha}{\epsilon} \frac{\Lambda_\mathrm{tot}}{2 u}, 
\end{equation}
where $u \equiv \Lambda/\Lambda_\mathrm{tot}$ is the expected lensing fraction. Note that $\kappa$ is a function of cosmological parameters and the observing period $\Tobs$ as the lensing fraction depends on $\vOmega$ and the total number of observed events depends on $\Tobs$. For simplicity of notation, we don't explicitly write down its dependence on $\vOmega$ and $\Tobs$. The contamination fraction also depends on $k_0 \equiv \alpha/\epsilon$, which depends on the receiver operating characteristic (ROC) of a given lensing identification method. This can be estimated by performing the same analysis on simulated lensed and unlensed GW events (see, e.g.~\cite{haris2018identifying}). To keep the contamination fraction low (about $10\%$), we would need $\alpha \sim 10^{-9}$ for $\Lambda \sim10^6$ and $u=0.01$. This is likely to be achievable in future GW observations due to the increased precision of measurements. 

It is easy to see that, with a contamination fraction $\kappa$, the expected number of lensed events will change to 
\begin{equation}
	\Lambda_{\mathrm{c}}(\vOmega, \Tobs) = \epsilon \left[1+\kappa \right]  \Lambda(\vOmega, \Tobs). 
\end{equation}
Thus, once $k_0$ is known from simulations, it is possible to model the effect of contamination on the expected number of lensed events. Similarly, the time delay distribution of the detected events will be a mixture of the lensed and unlensed time delays. We model the effect of contamination on the time delay distribution as:
\begin{equation}
	p_\mathrm{c}(\Delta t~|~\vec{\Omega}, \Tobs) = \frac{\kappa}{1+\kappa} p_{\mathrm{unlens}}(\Delta t ~|~ \Tobs) + \frac{1}{1+\kappa} ~ p_\mathrm{lens}(\Delta t~|~\vOmega, \Tobs), 
\end{equation}
and use them for the cosmological parameter estimation. Above, $p_\mathrm{lens}(\Delta t~|~\vOmega, \Tobs)$ is given by equation~(\ref{eq:P_DeltaT_obs_lens}), while 
\begin{equation}
p_{\mathrm{unlens}}(\Delta t ~|~ \Tobs) \propto (\Tobs- \Delta t) \Theta(\Tobs- \Delta t). 
\end{equation}

\begin{figure}[tbh]
	\centering 
	\includegraphics[width=0.40\textwidth]{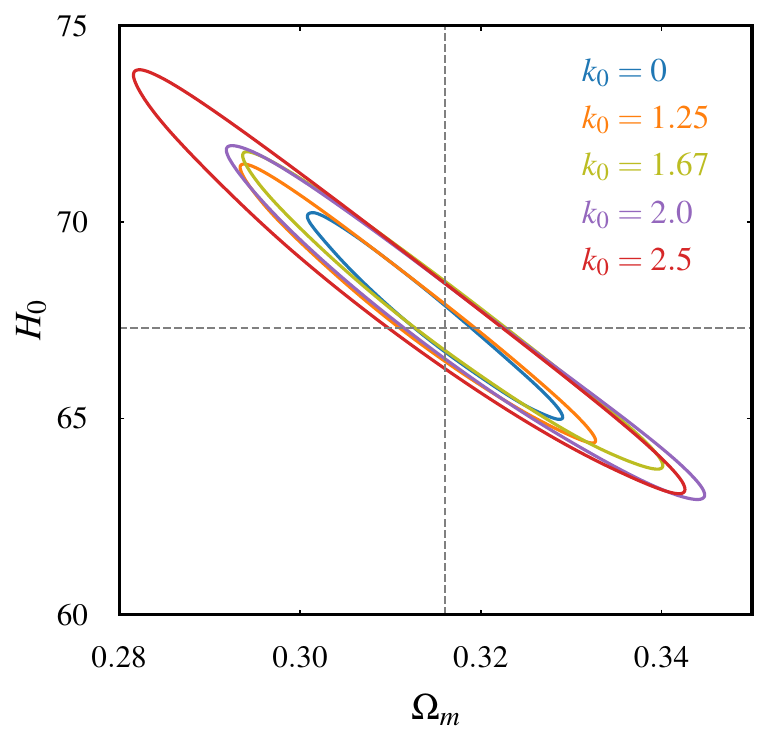}
	\includegraphics[width=0.4158\columnwidth]{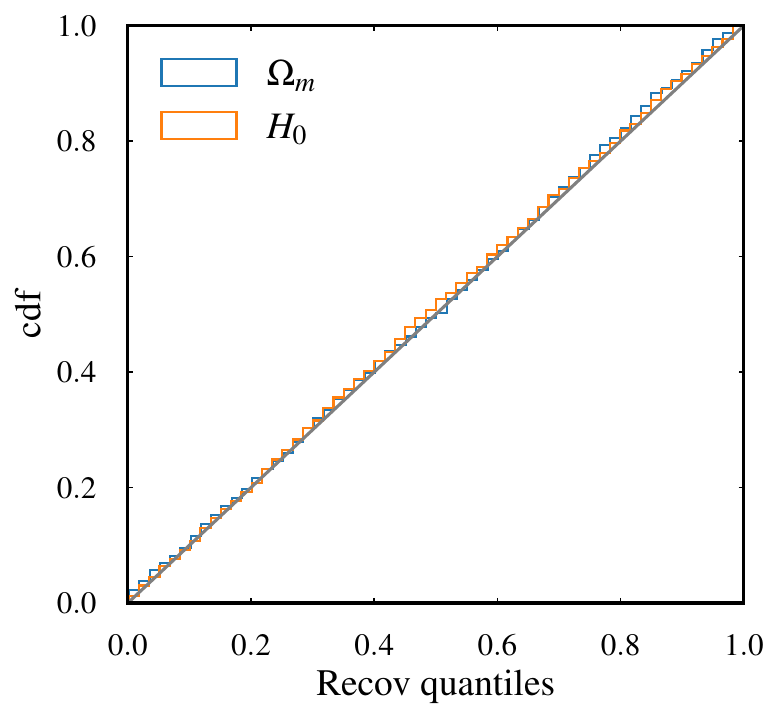}
	\caption{\emph{Left panel:} Posteriors on the cosmological parameters (95\% credible regions) when we assume different amounts of data contamination. The legends show $k_0 \equiv \alpha/\epsilon$ in units of $10^{-9}$, where $\alpha$ ($\epsilon$) is the false (true) positive probability of the lensing identification algorithm. Data contamination only worsens the precision of the posteriors, without causing any systematic biases. {\emph{Right panel:} p-p plot obtained from $\sim 10^3$ recovery tests for $\alpha=10^{-9}$ and $\epsilon=0.5$. This indicates no systematic bias in cosmological parameter recovery when we account for contamination.} }
	\label{fig:posterir-contaminated}
\end{figure}

To gauge the effect of contamination on the cosmological parameter inference, we simulate lensed events following the {Dominick} redshift distribution, assuming the true cosmology $\Omega_{\mathrm{true}} = \{\Omega_m = 0.316, H_0 = 67.3\}$, with $R=5\times10^5$ and $T_{\mathrm{obs}} = 10\; \mathrm{yrs}$. This corresponds to a true expected detection of $\Lambda_\mathrm{tot} = 5\times10^6$ binaries. We assume that the lensing identification method has a false positive probability of $\alpha=10^{-9}$. We assume different true positive probabilities:  $\epsilon = 0.4, 0.5, 0.6, 0.8$, corresponding to $k_0 \equiv \alpha/\epsilon = [2.5, 2, 1.67, 1.25] \times 10^{-9}$. Figure \ref{fig:posterir-contaminated} shows the posteriors on the cosmological parameters from an analysis that takes into the effect of contamination. When there is no contamination $k_0 = 0$, we reproduce the results from figure~\ref{fig:zdist}. Additionally, data contamination only worsens the precision of our measurement, without causing any systematic biases ({see p-p plot in figure~\ref{fig:posterir-contaminated}}). Table~\ref{tab:contaminated} tabulates the expected precision in the measurement of cosmological parameters with different levels of contamination. 

\begin{table}[h!]
	\centering
	\begin{tabular}{cccccc} 
		\hline
		\hline 
		& No contamination & $k_0 = 2.5$ & $k_0 = 2$ & $k_0 = 1.67$& $k_0 = 1.25$\\ 
		\hline
		\hline 
		$\Lambda_{\mathrm{c}}\left(\Omega_{\mathrm{true}}\right)$& 37700 & 27580 & 31350 & 35120& 42660\\  
		$\Omega_m(68\%)$& $0.315^{+0.006}_{-0.006}$ & $0.311^{+0.012}_{-0.012}$ & $0.317^{+0.011}_{-0.011}$ &$0.317^{+0.009}_{-0.009}$ &$0.312^{+0.008}_{-0.008}$\\ 
		$\Omega_m(95\%)$& $0.315^{+0.012}_{-0.011}$ & $0.311^{+0.025}_{-0.023}$ & $0.317^{+0.022}_{-0.020}$ &$0.317^{+0.019}_{-0.018}$ &$0.312^{+0.016}_{-0.015}$\\ 
		$H_0(68\%)$& $67.6^{+1.1}_{-1.1}$ & $68.3^{+2.2}_{-2.2}$ & $67.3^{+1.8}_{-1.8}$ &$67.6^{+1.6}_{-1.6}$ & $67.9^{+1.4}_{-1.4}$ \\ 
		$H_0(95\%)$& $67.6^{+2.1}_{-2.0}$ & $68.3^{+4.4}_{-4.3}$ & $67.3^{+3.7}_{-3.5}$ &$67.6^{+3.3}_{-3.1}$ & $67.9^{+2.8}_{-2.8}$ \\ 
		\hline
		\hline
	\end{tabular}
	\caption{Expected number $\Lambda_{\mathrm{c}}$  of lensed events after considering contamination for $\vec{\Omega}=\vec{\Omega}_{\mathrm{true}}$ and expected constraints (68\% and 95\% credible intervals) on $\Omega_m$ and $H_0$ for different values of the true positive probability $\epsilon$ for the the data analysis method that is used to identify lensed events (assuming a false positive probability $\alpha = 10^{-9}$). We assume a  merger rate of $5\times10^5\;\mathrm{yr}^{-1}$ and an observation time of $10\;\mathrm{yrs}$.}
	\label{tab:contaminated}
\end{table}

\section{Conclusions and future work}

In this paper we presented a detailed exposition of a statistical method for cosmography from the observation of a large number ($\sim 10^4$) of strongly lensed BBH mergers observable by XG detectors. This method, first presented in~\cite{Jana_2023}, compares the observed number of strongly lensed GW events and their time delay distribution (between lensed images) with observed events to infer cosmological parameters. We showed that the prospective constraints from the XG detectors are comparable to the current best measurements, but probing a cosmological epoch ($z \sim 1-10$) that is not probed by other observations. This can potentially shed some light on the ``tensions'' that exist between current cosmological observations~\cite{ABDALLA202249}, some hinting at the failure of the flat $\Lambda$CDM model. 

The observed number of lensed events and their time delay distribution will depend on the distribution of GW sources as well as lenses, apart from the cosmological parameters. Properties of the GW source distribution could be measured accurately from the large number ($\sim 10^6$) of unlensed GW signals that will dominate the data. Since the distribution of the GW source properties is currently largely unknown, we considered a few different astrophysical models for the same, and showed that the expected constraints don't vary significantly. {We also showed that the statistical uncertainties in the reconstruction of source properties from GW observations will not significantly affect the inference of cosmological parameters, as long as the source redshift distribution can be reconstructed in an unbiased manner.} We also showed that the limited ability of our data analysis algorithms to distinguish between lensed and unlensed GW events, resulting in some amount of contamination in the sample of lensed GW events, will not bias our inference. This can be avoided by incorporating the effect of data contamination in our Bayesian likelihood models. 

One of the major sources of error in our analysis is likely to come from the uncertainty in the distribution of gravitational lenses. In this paper, we assumed that the properties of the lens distribution can be extracted from theoretical HMF models. We used several HMFs that model the expected mass distribution of dark matter halos at different redshifts. We then assumed that these halos are spherical symmetric and used a simple prescription to map the mass of the dark matter halo to the velocity dispersion of the lens in the SIS model. If the HMF model that we use is significantly different from the ``true'' model, that can bias the estimation of cosmological parameters. However, in such situations, Bayesian model selection involving several HMF models should enable us to identify the ``true'' model of the HMF. In the future, we could improve this method by employing parametric models of HMFs that include modelling errors, which can be marginalised over in the Bayesian inference. A combination of EM observations and cosmological simulations can be used to further improve our models of the distribution of lens properties. 

The SIS model that we use for lenses is also an oversimplified model. In a follow-up work we plan to incorporate more realistic lens models in our analysis and better prescriptions for mapping the mass distribution of dark matter halos to the lens properties. Our ongoing investigations suggest that the changes in the mass distribution of dark matter halos is not degenerate with changes in cosmological parameters. This suggests that the same method could also be used to probe the mass distribution of dark matter halos at various redshifts, thus effectively probing the structure formation and the nature of dark matter. This is explored in our ongoing work. 

While we focus on BBH mergers in this paper, observations of strongly lensed binaries involving neutron stars could also be powerful.  Since BNS could be observed only out to a smaller redshifts ($z \lesssim 2$ even for XG detectors), their lensing optical depth will be smaller compared to BBHs which could be observed out to very large redshifts ($ z \sim 10-100$). However, BNS mergers are expected to be large in number ($\sim 10^6$ detections per year in XG~\cite{Baibhav_2019}) and, owing to their long signal duration in the detector band, their properties could be measured with much better accuracy. Furthermore, some of them could also produce observable EM counterparts. Strongly lensed BNSs could be very powerful in probing low-redshift cosmology. We are exploring this in an ongoing work.  In conclusion, future observations of strongly lensed GWs will enable several powerful new probes of cosmology. 

\subsection*{Acknowledgments:} We thank Otto Akseli Hannuksela for his careful review of the manuscript and useful comments. We are also thankful to Ken K.Y. Ng and Salvatore Vitale for supplying the posterior predicted samples of merger rate density, which were utilised in the analysis conducted in section \ref{sec:measurement-error}. For useful discussions, we thank Koustav Narayan Maity as well as members of the ICTS Astrophysical Relativity group and the LIGO-Virgo-KAGRA collaboration's working group on gravitational lensing. Our research is supported by the Department of Atomic Energy, Government of India, under Project No. RTI4001. The numerical calculations were carried out using the Alice cluster at ICTS.

\section*{References}
\bibliographystyle{iopart-num}
\bibliography{references.bib}

\providecommand{\newblock}{}
\begin{thebibliography}{10}
\expandafter\ifx\csname url\endcsname\relax
  \def\url#1{{\tt #1}}\fi
\expandafter\ifx\csname urlprefix\endcsname\relax\def\urlprefix{URL }\fi
\providecommand{\eprint}[2][]{\url{#2}}

\bibitem{Maggiore_2020}
Maggiore M, Broeck C~V~D, Bartolo N {\em et~al.\/} 2020 {\em Journal of
  Cosmology and Astroparticle Physics\/} {\bf 2020} 050
  \urlprefix\url{https://dx.doi.org/10.1088/1475-7516/2020/03/050}

\bibitem{evans2023ce}
Evans M, Corsi A, Afle C, Ananyeva A, Arun K~G {\em et~al.\/} 2023 Cosmic
  explorer: A submission to the nsf mpsac nggw subcommittee (\textit{Preprint}
  \eprint{2306.13745})

\bibitem{Hall_2019}
Hall E~D and Evans M 2019 {\em Classical and Quantum Gravity\/} {\bf 36} 225002
  \urlprefix\url{https://doi.org/10.1088%2F1361-6382%2Fab41d6}

\bibitem{SchutzNat1986}
Schutz B~F 1986 {\em Nature\/} {\bf 323} 310--311 ISSN 1476-4687
  \urlprefix\url{https://doi.org/10.1038/323310a0}

\bibitem{Holz_Hughes_2005}
Holz D~E and Hughes S~A 2005 {\em The Astrophysical Journal\/} {\bf 629} 15
  \urlprefix\url{https://dx.doi.org/10.1086/431341}

\bibitem{Riess2020}
Riess A~G 2020 {\em Nature Reviews Physics\/} {\bf 2} 10--12 ISSN 2522-5820
  \urlprefix\url{https://doi.org/10.1038/s42254-019-0137-0}

\bibitem{Hu_2023}
Hu J~P and Wang F~Y 2023 {\em Universe\/} {\bf 9} ISSN 2218-1997
  \urlprefix\url{https://www.mdpi.com/2218-1997/9/2/94}

\bibitem{Fishbach_2019}
Fishbach M, Gray R, Hernandez I~M {\em et~al.\/} 2019 {\em The Astrophysical
  Journal Letters\/} {\bf 871} L13
  \urlprefix\url{https://dx.doi.org/10.3847/2041-8213/aaf96e}

\bibitem{Calore_2020}
Calore F, Cuoco A, Regimbau T, Sachdev S and Serpico P~D 2020 {\em Phys. Rev.
  Res.\/} {\bf 2}(2) 023314
  \urlprefix\url{https://link.aps.org/doi/10.1103/PhysRevResearch.2.023314}

\bibitem{mukherjee2022crosscorrelating}
Mukherjee S, Krolewski A, Wandelt B~D and Silk J 2022 Cross-correlating dark
  sirens and galaxies: measurement of $h_0$ from gwtc-3 of ligo-virgo-kagra
  (\textit{Preprint} \eprint{2203.03643})

\bibitem{Gair_2023}
Gair J~R, Ghosh A, Gray R, Holz D~E, Mastrogiovanni S, Mukherjee S, Palmese A,
  Tamanini N, Baker T, Beirnaert F, Bilicki M, Chen H~Y, Dálya G, Ezquiaga
  J~M, Farr W~M, Fishbach M, Garcia-Bellido J, Ghosh T, Huang H~Y, Karathanasis
  C, Leyde K, Hernandez I~M, Noller J, Pierra G, Raffai P, Romano A~E,
  Seglar-Arroyo M, Steer D~A, Turski C, Vaccaro M~P and Vallejo-Peña S~A 2023
  {\em The Astronomical Journal\/} {\bf 166} 22
  \urlprefix\url{https://dx.doi.org/10.3847/1538-3881/acca78}

\bibitem{Borghi_2024}
Borghi N, Mancarella M, Moresco M, Tagliazucchi M, Iacovelli F, Cimatti A and
  Maggiore M 2024 {\em The Astrophysical Journal\/} {\bf 964} 191 ISSN
  1538-4357 \urlprefix\url{http://dx.doi.org/10.3847/1538-4357/ad20eb}

\bibitem{Planck_2018}
{Planck Collaboration}, {Aghanim, N}, {Akrami, Y}, {Ashdown, M}, {Aumont, J}
  and {others} 2020 {\em A\&A\/} {\bf 641} A6
  \urlprefix\url{https://doi.org/10.1051/0004-6361/201833910}

\bibitem{Lattimer_2021}
Lattimer J 2021 {\em Annual Review of Nuclear and Particle Science\/} {\bf 71}
  433--464 ISSN 1545-4134
  \urlprefix\url{https://www.annualreviews.org/content/journals/10.1146/annurev-nucl-102419-124827}

\bibitem{Woosley2007}
Woosley S~E, Blinnikov S and Heger A 2007 {\em Nature\/} {\bf 450} 390--392
  ISSN 1476-4687 \urlprefix\url{https://doi.org/10.1038/nature06333}

\bibitem{Belczynski_2016a}
{Belczynski, K}, {Heger, A}, {Gladysz, W}, {Ruiter, A J}, {Woosley, S},
  {Wiktorowicz, G}, {Chen, H-Y}, {Bulik, T}, {O’Shaughnessy, R}, {Holz, D E},
  {Fryer, C L} and {Berti, E} 2016 {\em A\&A\/} {\bf 594} A97
  \urlprefix\url{https://doi.org/10.1051/0004-6361/201628980}

\bibitem{Woosley_2017}
Woosley S~E 2017 {\em The Astrophysical Journal\/} {\bf 836} 244
  \urlprefix\url{https://dx.doi.org/10.3847/1538-4357/836/2/244}

\bibitem{Heger_2002}
Heger A and Woosley S~E 2002 {\em The Astrophysical Journal\/} {\bf 567} 532
  \urlprefix\url{https://dx.doi.org/10.1086/338487}

\bibitem{Mastrogiovanni:2021wsd}
Mastrogiovanni S, Leyde K, Karathanasis C, Chassande-Mottin E, Steer D~A, Gair
  J, Ghosh A, Gray R, Mukherjee S and Rinaldi S 2021 {\em Phys. Rev. D\/} {\bf
  104} 062009 (\textit{Preprint} \eprint{2103.14663})

\bibitem{Mukherjee:2021rtw}
Mukherjee S 2022 {\em Mon. Not. Roy. Astron. Soc.\/} {\bf 515} 5495--5505
  (\textit{Preprint} \eprint{2112.10256})

\bibitem{Messenger:2011gi}
Messenger C and Read J 2012 {\em Phys. Rev. Lett.\/} {\bf 108} 091101
  (\textit{Preprint} \eprint{1107.5725})

\bibitem{Li:2013via}
Li T~G~F, Del~Pozzo W and Messenger C 2015 {\em {13th Marcel Grossmann Meeting
  on Recent Developments in Theoretical and Experimental General Relativity,
  Astrophysics, and Relativistic Field Theories}\/} pp 2019--2021
  (\textit{Preprint} \eprint{1303.0855})

\bibitem{Messenger:2013fya}
Messenger C, Takami K, Gossan S, Rezzolla L and Sathyaprakash B~S 2014 {\em
  Phys. Rev. X\/} {\bf 4} 041004 (\textit{Preprint} \eprint{1312.1862})

\bibitem{Ezquiaga_2021_PISN}
Ezquiaga J~M and Holz D~E 2021 {\em The Astrophysical Journal Letters\/} {\bf
  909} L23 ISSN 2041-8213
  \urlprefix\url{http://dx.doi.org/10.3847/2041-8213/abe638}

\bibitem{Ezquiaga_2022}
Ezquiaga J~M and Holz D~E 2022 {\em Physical Review Letters\/} {\bf 129} ISSN
  1079-7114 \urlprefix\url{http://dx.doi.org/10.1103/PhysRevLett.129.061102}

\bibitem{chen2024cosmography}
Chen H~Y, Ezquiaga J~M and Gupta I 2024 Cosmography with next-generation
  gravitational wave detectors (\textit{Preprint} \eprint{2402.03120})

\bibitem{Belgacem:2018lbp}
Belgacem E, Dirian Y, Foffa S and Maggiore M 2018 {\em Phys. Rev. D\/} {\bf 98}
  023510 (\textit{Preprint} \eprint{1805.08731})

\bibitem{Mancarella_2022}
Mancarella M, Genoud-Prachex E and Maggiore M 2022 {\em Phys. Rev. D\/} {\bf
  105}(6) 064030
  \urlprefix\url{https://link.aps.org/doi/10.1103/PhysRevD.105.064030}

\bibitem{Vijaykumar_2023}
Vijaykumar A, Saketh M~V~S, Kumar S, Ajith P and Choudhury T~R 2023 {\em Phys.
  Rev. D\/} {\bf 108}(10) 103017
  \urlprefix\url{https://link.aps.org/doi/10.1103/PhysRevD.108.103017}

\bibitem{Kumar_2022}
Kumar S, Vijaykumar A and Nitz A~H 2022 {\em The Astrophysical Journal\/} {\bf
  930} 113 \urlprefix\url{https://dx.doi.org/10.3847/1538-4357/ac5e34}

\bibitem{Li2018}
{Li} S~S, {Mao} S, {Zhao} Y and {Lu} Y 2018 {\em MNRAS\/} {\bf 476} 2220--2229
  (\textit{Preprint} \eprint{1802.05089})

\bibitem{Ng2018}
Ng K~K~Y, Wong K~W~K, Broadhurst T and Li T~G~F 2018 {\em Phys. Rev. D\/} {\bf
  97}(2) 023012
  \urlprefix\url{https://link.aps.org/doi/10.1103/PhysRevD.97.023012}

\bibitem{Jana_2023}
Jana S, Kapadia S~J, Venumadhav T and Ajith P 2023 {\em Phys. Rev. Lett.\/}
  {\bf 130}(26) 261401
  \urlprefix\url{https://link.aps.org/doi/10.1103/PhysRevLett.130.261401}

\bibitem{Xu_2022}
Xu F, Ezquiaga J~M and Holz D~E 2022 {\em The Astrophysical Journal\/} {\bf
  929} 9 \urlprefix\url{https://dx.doi.org/10.3847/1538-4357/ac58f8}

\bibitem{Yang_2021}
Yang L, Wu S, Liao K, Ding X, You Z, Cao Z, Biesiada M and Zhu Z~H 2021 {\em
  Monthly Notices of the Royal Astronomical Society\/} {\bf 509} 3772--3778
  ISSN 0035-8711 (\textit{Preprint}
  \eprint{https://academic.oup.com/mnras/article-pdf/509/3/3772/41386274/stab3298.pdf})
  \urlprefix\url{https://doi.org/10.1093/mnras/stab3298}

\bibitem{gupta2024characterizing}
Gupta I, Afle C, Arun K~G, Bandopadhyay A {\em et~al.\/} 2024 Characterizing
  gravitational wave detector networks: From a$^\sharp$ to cosmic explorer
  (\textit{Preprint} \eprint{2307.10421})

\bibitem{Oguri_2018}
Oguri M 2018 {\em Monthly Notices of the Royal Astronomical Society\/} {\bf
  480} 3842--3855 ISSN 0035-8711 (\textit{Preprint}
  \eprint{https://academic.oup.com/mnras/article-pdf/480/3/3842/25520286/sty2145.pdf})
  \urlprefix\url{https://doi.org/10.1093/mnras/sty2145}

\bibitem{Smith_2023}
Smith G~P, Robertson A, Mahler G {\em et~al.\/} 2023 {\em Monthly Notices of
  the Royal Astronomical Society\/} {\bf 520} 702--721 ISSN 0035-8711
  (\textit{Preprint}
  \eprint{https://academic.oup.com/mnras/article-pdf/520/1/702/49032005/stad140.pdf})
  \urlprefix\url{https://doi.org/10.1093/mnras/stad140}

\bibitem{Wierda_2021}
Wierda A~R~A~C, Wempe E, Hannuksela O~A, Koopmans L~V~E and Van Den~Broeck C
  2021 {\em The Astrophysical Journal\/} {\bf 921} 154 ISSN 1538-4357
  \urlprefix\url{http://dx.doi.org/10.3847/1538-4357/ac1bb4}

\bibitem{haris2018identifying}
Haris K, Mehta A~K, Kumar S, Venumadhav T and Ajith P 2018 Identifying strongly
  lensed gravitational wave signals from binary black hole mergers
  (\textit{Preprint} \eprint{1807.07062})

\bibitem{wempe2022lensing}
Wempe E, Koopmans L~V~E, Wierda A~R~A~C, Hannuksela O~A and van~den Broeck C
  2022 A lensing multi-messenger channel: Combining ligo-virgo-kagra lensed
  gravitational-wave measurements with euclid observations (\textit{Preprint}
  \eprint{2204.08732})

\bibitem{Vitale_2019}
Vitale S, Farr W~M, Ng K~K~Y and Rodriguez C~L 2019 {\em The Astrophysical
  Journal Letters\/} {\bf 886} L1
  \urlprefix\url{https://dx.doi.org/10.3847/2041-8213/ab50c0}

\bibitem{Liao:2017ioi}
Liao K, Fan X~L, Ding X~H, Biesiada M and Zhu Z~H 2017 {\em Nature Commun.\/}
  {\bf 8} 1148 [Erratum: Nature Commun. 8, 2136 (2017)] (\textit{Preprint}
  \eprint{1703.04151})

\bibitem{LISA_cosmography}
Sereno M, Jetzer P, Sesana A and Volonteri M 2011 {\em Monthly Notices of the
  Royal Astronomical Society\/} {\bf 415} 2773--2781 ISSN 0035-8711
  (\textit{Preprint}
  \eprint{https://academic.oup.com/mnras/article-pdf/415/3/2773/5980940/mnras0415-2773.pdf})
  \urlprefix\url{https://doi.org/10.1111/j.1365-2966.2011.18895.x}

\bibitem{Otto_localise_h0}
Hannuksela O~A, Collett T~E, Çalışkan M and Li T~G~F 2020 {\em Monthly
  Notices of the Royal Astronomical Society\/} {\bf 498} 3395--3402 ISSN
  0035-8711 (\textit{Preprint}
  \eprint{https://academic.oup.com/mnras/article-pdf/498/3/3395/33779471/staa2577.pdf})
  \urlprefix\url{https://doi.org/10.1093/mnras/staa2577}

\bibitem{shan2023microlensing}
Shan X, Chen X, Hu B and Cai R~G 2023 Microlensing sheds light on the detection
  of strong lensing gravitational waves (\textit{Preprint} \eprint{2301.06117})

\bibitem{Dominik_2013}
Dominik M, Belczynski K, Fryer C, Holz D~E, Berti E, Bulik T, Mandel I and
  O'Shaughnessy R 2013 {\em The Astrophysical Journal\/} {\bf 779} 72
  \urlprefix\url{https://dx.doi.org/10.1088/0004-637X/779/1/72}

\bibitem{Oguri_2002}
Oguri M, Taruya A, Suto Y and Turner E~L 2002 {\em The Astrophysical Journal\/}
  {\bf 568} 488 \urlprefix\url{https://dx.doi.org/10.1086/339064}

\bibitem{Belczynski_2016nat}
Belczynski K, Holz D~E, Bulik T and O'Shaughnessy R 2016 {\em Nature\/} {\bf
  534} 512--515 ISSN 1476-4687
  \urlprefix\url{https://doi.org/10.1038/nature18322}

\bibitem{Madau_Dickinson_2014}
Madau P and Dickinson M 2014 {\em Annual Review of Astronomy and
  Astrophysics\/} {\bf 52} 415--486 (\textit{Preprint}
  \eprint{https://doi.org/10.1146/annurev-astro-081811-125615})
  \urlprefix\url{https://doi.org/10.1146/annurev-astro-081811-125615}

\bibitem{Behroozi_2013}
Behroozi P~S, Wechsler R~H and Conroy C 2013 {\em The Astrophysical Journal\/}
  {\bf 770} 57 \urlprefix\url{https://dx.doi.org/10.1088/0004-637X/770/1/57}

\bibitem{murray2013hmfcalc}
Murray S, Power C and Robotham A 2013 Hmfcalc: An online tool for calculating
  dark matter halo mass functions (\textit{Preprint} \eprint{1306.6721})

\bibitem{Rinaldi:2021bhm}
Rinaldi S and Del~Pozzo W 2021 {\em Mon. Not. Roy. Astron. Soc.\/} {\bf 509}
  5454--5466 (\textit{Preprint} \eprint{2109.05960})

\bibitem{Jenkins_2001}
Jenkins A, Frenk C~S, White S~D~M, Colberg J~M, Cole S, Evrard A~E, Couchman
  H~M~P and Yoshida N 2001 {\em Monthly Notices of the Royal Astronomical
  Society\/} {\bf 321} 372--384 ISSN 0035-8711 (\textit{Preprint}
  \eprint{https://academic.oup.com/mnras/article-pdf/321/2/372/3032036/321-2-372.pdf})
  \urlprefix\url{https://doi.org/10.1046/j.1365-8711.2001.04029.x}

\bibitem{SMT_2001}
Sheth R~K, Mo H~J and Tormen G 2001 {\em Monthly Notices of the Royal
  Astronomical Society\/} {\bf 323} 1--12 ISSN 0035-8711 (\textit{Preprint}
  \eprint{https://academic.oup.com/mnras/article-pdf/323/1/1/3204200/323-1-1.pdf})
  \urlprefix\url{https://doi.org/10.1046/j.1365-8711.2001.04006.x}

\bibitem{Tinker_2008}
Tinker J, Kravtsov A~V, Klypin A, Abazajian K, Warren M, Yepes G, Gottlöber S
  and Holz D~E 2008 {\em The Astrophysical Journal\/} {\bf 688} 709
  \urlprefix\url{https://dx.doi.org/10.1086/591439}

\bibitem{Gorenstein_1988}
{Gorenstein} M~V, {Falco} E and {Shapiro} I 1988 {\em \apj\/} {\bf 327} 693

\bibitem{Saha_2000}
Saha P 2000 {\em The Astronomical Journal\/} {\bf 120} 1654
  \urlprefix\url{https://dx.doi.org/10.1086/301581}

\bibitem{Falco_1985}
{Falco} E, {Gorenstein} M and {Shapiro} I 1985 {\em \apjl\/} {\bf 289} L1--L4

\bibitem{Kochanek_2002}
Kochanek C~S 2002 {\em The Astrophysical Journal\/} {\bf 578} 25
  \urlprefix\url{https://dx.doi.org/10.1086/342476}

\bibitem{Poon:2024}
Poon J~S, Rinaldi S, Janquart J, Narola H and Hannuksela O~A
  (\textit{Preprint} \eprint{{LIGO-P2400173}})

\bibitem{Janquart_2023}
Janquart J, More A and Van Den Broeck C 2022 {\em Monthly Notices of the
  Royal Astronomical Society\/} {\bf 519} 2046--2059 ISSN 0035-8711
  (\textit{Preprint}
  \eprint{https://academic.oup.com/mnras/article-pdf/519/2/2046/48451053/stac3660.pdf})
  \urlprefix\url{https://doi.org/10.1093/mnras/stac3660}

\bibitem{Caliskan_lensing_luck_2023}
\ifmmode \mbox{\c{C}}\else \c{C}\fi{}al\ifmmode \imath \else \i
  \fi{}\ifmmode~\mbox{\c{s}}\else \c{s}\fi{}kan M, Ezquiaga J~M, Hannuksela O~A
  and Holz D~E 2023 {\em Phys. Rev. D\/} {\bf 107}(6) 063023
  \urlprefix\url{https://link.aps.org/doi/10.1103/PhysRevD.107.063023}

\bibitem{ABDALLA202249}
Abdalla E, Abellán G~F, Aboubrahim A {\em et~al.\/} 2022 {\em Journal of High
  Energy Astrophysics\/} {\bf 34} 49--211 ISSN 2214-4048
  \urlprefix\url{https://www.sciencedirect.com/science/article/pii/S2214404822000179}

\bibitem{Baibhav_2019}
Baibhav V, Berti E, Gerosa D, Mapelli M, Giacobbo N, Bouffanais Y and Di~Carlo
  U~N 2019 {\em Phys. Rev. D\/} {\bf 100}(6) 064060
  \urlprefix\url{https://link.aps.org/doi/10.1103/PhysRevD.100.064060}

\end{thebibliography}
\end{document}